\documentclass[journal]{IEEEtran}

\usepackage[T1]{fontenc}
\usepackage{color, colortbl}
\usepackage{babel}
\usepackage{verbatim}
\usepackage{textcomp}
\usepackage{amsmath}
\usepackage{amsthm}
\usepackage{amssymb}
\usepackage{amstext}
\usepackage{graphicx}
\usepackage{tabulary}
\usepackage{caption}
\usepackage{subcaption}
\usepackage{epstopdf}
\usepackage{cite}
\usepackage[normalem]{ulem}
\usepackage{balance}
\usepackage{multirow}
\usepackage{xcolor}
\usepackage{soul}
\usepackage[ruled,vlined]{algorithm2e}
\addto\captionsenglish{}
\usepackage{bm}
\usepackage{soul}

\def\RSS{\mathtt{RSS}}

\def\SINR{\mathtt{SINR}}

\newlength \figwidth
\newtheorem{Proposition}{\bf Proposition}

\newtheorem{Remark}{\bf Remark}

\allowdisplaybreaks

\begin{document}

\title{Optimizing Cellular Networks for UAV Corridors via Quantization Theory}

\author{\IEEEauthorblockN{{Saeed Karimi-Bidhendi, Giovanni Geraci, and Hamid Jafarkhani}} 
\thanks{Saeed Karimi-Bidhendi and Hamid Jafarkhani are with the Center for Pervasive Communications \& Computing, University of California, Irvine, Irvine CA, 92697 USA (e-mail: \{skarimib, hamidj\}$@$uci.edu).}
\thanks{Giovanni~Geraci is with Telef\'{o}nica Research and Universitat Pompeu Fabra (UPF), Barcelona, Spain (e-mail: giovanni.geraci$@$upf.edu).}
\thanks{This work was supported in part by the NSF Award CNS-2229467, by the Spanish Research Agency through grants PID2021-123999OB-I00, CEX2021-001195-M, and the ``Ram\'{o}n y Cajal" program, and by the UPF-Fractus Chair on Tech Transfer and 6G. Some of the results in this paper were presented at IEEE ICC'23 \cite{KarGerJafICC2023}.}
}

\maketitle

\begin{abstract}


We present a new framework based on quantization theory to design cellular networks optimized for both legacy ground users and uncrewed aerial vehicle (UAV) corridors, dedicated aerial highways for safe UAV flights. Our framework leverages antenna tilts and transmit power at each base station to enhance coverage and quality of service among users. 
We develop a comprehensive mathematical analysis and optimization algorithms for multiple system-level performance metrics, including received signal strength and signal-to-interference-plus-noise ratio. Realistic antenna radiation patterns and propagation channel models are considered, alongside a generic 3D user distribution that allows for performance prioritization on the ground, along UAV corridors, or a desired tradeoff between the two. We demonstrate the efficacy of the proposed framework through case studies, showcasing the non-trivial combinations of antenna tilts and power levels that improve coverage and signal quality along UAV corridors while incurring only a marginal impact on the ground user performance compared to scenarios without UAVs.
\end{abstract}


\IEEEpeerreviewmaketitle

\section{Introduction}\label{Introduction}


\subsection{Motivation and Related Work}

Uncrewed aerial vehicles (UAVs), commonly known as drones, are expected to contribute to extraordinary economic growth and societal transformations. Thanks to their low cost and high mobility, UAVs will become of paramount importance for goods delivery, surveillance, search and rescue, and the monitoring of wildfire, crowds, and assets \cite{GerGarAza2022,wu20205g,UAVfire2022}. With rising urbanization pushing ground transportation to its limits, electrical vertical take-off and landing vehicles (eVTOLs) serving as air taxis or ambulances would take urban mobility to new heights, contributing to a faster, safer, and more interconnected transportation system. Autonomous levitating pods are no longer science fiction as projects and tests are underway, and they could redefine how we commute and, in turn, where we live and work \cite{SaeAlnAlo2020}. 
For these and other applications, UAVs will need to exchange an unprecedented amount of real-time data with the network, requiring ultra-reliable wireless connectivity. The latter must support safe UAV operations through low-latency control and mission-specific data payloads, persuading legislators to ease the current regulations on civilian pilotless flights and giving the green light to autonomous UAVs and the associated vertical markets \cite{FotQiaDin2019,ZenGuvZha2020,SaaBenMoz2020,NamChaKim17,CDHJGC20}.

Achieving fly-and-connect capabilities faces important hurdles. Traditional cellular base stations (BSs) are designed to optimize 2D connectivity on the ground. As a result, UAVs can only be reached by their upper antenna sidelobes, and their movement causes sharp signal fluctuations. In addition, UAVs flying above buildings receive and create line-of-sight (LoS) interfering signals from a plurality of BSs. This interference results in UAVs experiencing a degraded signal-to-interference-plus-noise ratio (SINR), which hinders the correct decoding of critical command and control messages \cite{GerGarGal2018,ZenLyuZha2019}. 
The mobile industry and academia have long joined forces to pursue \emph{3D connectivity}, i.e., also up in the air, by re-engineering the deployments originally built for the ground. Short-term solutions are being implemented to handle a few network-connected UAVs without compromising the performance of existing ground users, e.g., via time/frequency separation \cite{NguAmoWig2018,3GPP36777}. However, this approach becomes increasingly inefficient as the number of UAVs grows because it requires dedicated radio resources for each UAV. 
More advanced proposals for ubiquitous aerial connectivity rely on network densification \cite{GarGerLop2019,KanMezLoz2021,DanGarGer2020,CDALHJTWC22}, dedicated infrastructure for aerial services \cite{GerLopBen2022,KimKimRyu2022,MozLinHay2021}, or satellites complementing the ground network \cite{BenGerLop2022}. Nonetheless, these proposals may require costly hardware or signal processing upgrades and still face difficulties providing ubiquitous connectivity to a multitude of aerial devices \cite{GerGarAza2022}.

\begin{figure}
\centering
\includegraphics[width=\figwidth]{
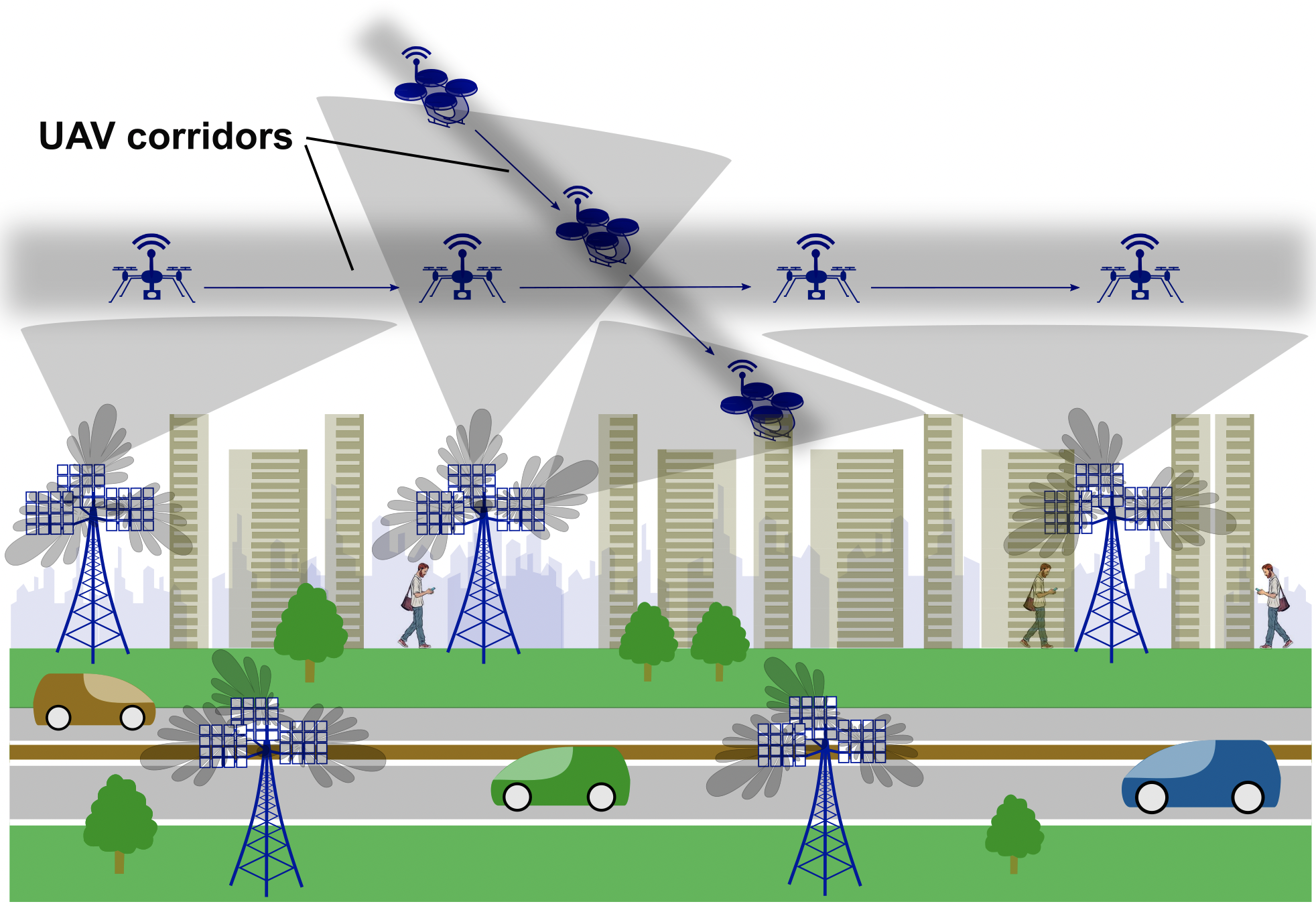}
\caption{Illustration of a cellular network with downtilted and uptilted BSs providing coverage to ground users as well as UAVs flying along corridors (blurred gray).}
\label{fig:illustration}
\end{figure}

The above circumstances rest on the assumptions that UAVs will fly unrestricted and cellular networks will need to guarantee connectivity at every 3D space location. However, just like ground vehicles and piloted aircrafts, as UAVs proliferate, their transit could be confined to specific aerial highways, denoted as \emph{UAV corridors} and defined by appropriate air traffic regulation authorities \cite{CheJaaYan2020,BhuGuvDai2021}. 
With the majority of UAVs flying along corridors, the operators' goal turns from providing sky-wide network services to guaranteeing corridor-wide reliable connectivity, as illustrated in Fig.~\ref{fig:illustration}. 
As the concept of UAV corridors gets traction, the community has been studying UAV trajectory optimization, matching the route of a UAV to the best network coverage pattern \cite{BulGuv2018,ChaSaaBet2018,EsrGanGes2020,BayTheCac2021}. However, the definition of UAV corridors will likely be network-agnostic and safety-driven, leaving very limited freedom for UAV trajectory optimization and a crucial need for a 3D network optimization. 
More recent work has targeted tuning cellular deployments to cater for UAV corridors through system-level simulations, large-scale optimization, or the theoretical analysis of a simplified setup \cite{MaeChoGuv2021,ChoGuvSaa2021,SinBhaOzt2021,BerLopGes2022,BerLopPio2023}. 
Nevertheless, there is an unmet need for a general mathematical framework allowing the analysis and design of cellular networks for both legacy ground users and UAV corridors. 
Stochastic geometry, a commonly used tool to analyze coverage and interference patterns for random spatial distributions of nodes (including UAVs \cite{azari2019cellular,AzaGerGar2020}) is not well-suited to optimize cellular networks for specific UAV corridors via deterministic BS deployments. 


\subsection{Contribution and Summary of Results}

In this paper, we take the first step towards creating such mathematical framework through quantization theory, already proven successful in addressing problems that involve the
geographical deployment of agents \cite{GuoJaf2016, cortes2005spatially, guo2018source, ingle2011energy, guo2019movement, cortes2004coverage, karimi2020energy, tang2019three, karimi2021energy, wang2006movement, 9086619,8519749}. Specifically, we determine the necessary conditions and design iterative algorithms to optimize the antenna tilts and transmit power at each BS of a cellular network to provide the best quality of service to both legacy ground users and UAVs flying along corridors. To the best of our knowledge, this is the first work doing so in a rigorous yet tractable manner, while accounting for a realistic network deployment, antenna radiation pattern, and propagation channel model. 

We conduct a comprehensive mathematical analysis and develop optimization algorithms for three system-level performance metrics, each averaged across all users within the target region: (i) average received signal strength (RSS), which serves as a proxy for coverage; (ii) average SINR, which serves as a proxy for quality of service; and (iii) max-product SINR and soft-max-min SINR, which allow to trade quality of service for fairness among users through tunable hyperparameters. Our analysis accommodates a generic 3D user distribution, enabling prioritization of performance on the ground, along UAV corridors, or any desired tradeoff between the two.

To illustrate the effectiveness of our mathematical framework, we further present multiple case studies, whose main takeaways can be summarized as follows:
\begin{itemize}
    \item 
    As expected, optimizing the antenna tilts for average RSS, with a focus on ground users or UAV corridors, results in all BSs either being downtilted or uptilted, respectively. However, by pursuing a tradeoff between the ground and the sky, we achieve a non-trivial combination of uptilted and downtilted antennas. This arrangement involves a subset of BSs catering to UAV corridors while maintaining coverage on the ground.
    \item
    Optimizing the network for SINR leads to a subset of BSs operating at maximum power, while the remaining ones operate at lower power levels or are altogether deactivated. This arrangement aims to provide a sufficiently strong signal while mitigating intercell interference, especially along UAV corridors.
    \item 
    Through the optimal combinations of antenna tilts and transmit power, which are non-obvious and otherwise difficult to design heuristically, our proposed algorithms significantly enhance coverage and signal quality along UAV corridors. These improvements are achieved with only a marginal reduction in ground performance compared to a scenario devoid of UAVs.
\end{itemize}

\section{System Model}\label{System-Model}

The cellular network under consideration is depicted in Fig.~\ref{fig:illustration} and detailed as follows.

\subsection{Network Topology}\label{Network-Topology}

\subsubsection{Ground Cellular Network} The underlying infrastructure of our network is a terrestrial cellular deployment consisting of $N$ BSs that provide service to network users. The height and $2$D location of BS $n$ is denoted by $h_{n,\mathrm{B}}$ and $\bm{p}_n$, respectively, for each $n \in \{1, \cdots, N\}$. Let $\bm{\Theta} = (\theta_1, \cdots, \theta_N)$ where $\theta_n \in [-90^\circ, +90^\circ]$ is the vertical antenna tilt of BS $n$, that can be adjusted by a mobile operator, with positive and negative angles denoting uptilts and downtilts, respectively. Let $\bm{\rho} = (\rho_1, \cdots, \rho_N)$ where $\rho_n$ is the transmission power of BS $n$, measured in dBm, which is also adjustable by a mobile operator with a maximum value of $\rho_{\max}$. We denote the antenna horizontal boresight direction (azimuth) of BS $n$ by $\phi_n \in [-180^\circ, +180^\circ]$ which is assumed to be fixed upon deployment.

\subsubsection{UAV Corridors and Legacy Ground Users} There are two types of users being served by the BSs: (i) UAVs that traverse a region $Q_U = \bigcup_{u=1}^{N_U} Q_u$ consisting of $N_U$ predefined 2D aerial routes/corridors $Q_u$; and (ii) ground-users (GUEs) that are dispersed over a 2D region $Q_G$. For each $1 \leq u \leq N_U$, all UAVs flying in the aerial corridor $Q_u$ are assumed to have a fixed height $h_u$. In addition, we consider a fixed height $h_G$ for all GUEs. Let $\lambda(\bm{q})$ be a probability density function that represents the distribution of users in the target region $Q = Q_U \bigcup Q_G$. Each user is associated with one BS; thus, the target region $Q$ is partitioned into $N$ disjoint subregions $\bm{V} = (V_1, \cdots, V_N)$ such that users within $V_n$ are associated with BS $n$.

\subsection{Channel Model}\label{Channel-Model}

\subsubsection{Antenna Gain} The BSs use  directional antennas with vertical and horizontal half-power beamwidths of $\theta_{\textrm{3dB}}$ and $\phi_{\textrm{3dB}}$, respectively. Let $A_{\max}$ be the maximum antenna gain at the boresight and denote the vertical and horizontal antenna gains in dB by $A_{n,\bm{q}}^{V}$ and $A_{n,\bm{q}}^{H}$, respectively. Directional antenna gains are given by \cite{3GPP38901}:
\begin{equation}\label{directional-antenna-gains}
    A_{n,\bm{q}}^{\mathrm{V}} =
    - \frac{12}{\theta^2_{\text{3dB}}} \left[ \theta_{n,\bm{q}} - \theta_n \right]^2, \quad
    A_{n,\bm{q}}^{\mathrm{H}} =
    - \frac{12}{\phi^2_{\text{3dB}}} \left[ \phi_{n,\bm{q}} - \phi_n \right]^2, 
\end{equation}
where $\theta_{n,\bm{q}}$ and $\phi_{n,\bm{q}}$ are the elevation angle and the azimuth angle between BS $n$ and the user location $\bm{q} \in Q$, respectively. These angles can be calculated as:
\begin{align}\label{theta-phi-nq}
    \theta_{n,\bm{q}} &= \tan^{-1}\!\left( \frac{h_{\bm{q}} - h_{n,\mathrm{B}}}{\| \bm{q} - \bm{p}_n \|}  \right), \\
    \phi_{n,\bm{q}} &= \!
\begin{cases}
\tan^{-1}\!\left(\frac{q_{\mathrm{y}}-p_{n,\mathrm{y}}}{q_{\mathrm{x}}-p_{n,\mathrm{x}}}\right) \!+\! 180^{\circ}\!\times\! 2c &  \!\!\!\text{if $q_{\mathrm{x}}-p_{n,\mathrm{x}}>0$}\\
\tan^{-1}\!\left(\frac{q_{\mathrm{y}}-p_{n,\mathrm{y}}}{q_{\mathrm{x}}-p_{n,\mathrm{x}}}\right) \!+\! 180^{\circ}\!\times\! (2c + 1) &  \!\!\!\text{if $q_{\mathrm{x}}-p_{n,\mathrm{x}}<0$}
\end{cases}\nonumber
\end{align}
where subscripts $\cdot_{\mathrm{x}}$ and $\cdot_{\mathrm{y}}$ denote the horizontal and vertical Cartesian coordinates of a point, respectively, and the integer $c$ is selected such that $-180^{\circ} \leq \phi_{n,\bm{q}} - \phi_n \leq +180^{\circ}$. Thus, the total antenna gain of BS $n$ in dB is given by $A_{n, \bm{q}} = A_{\max} + A_{n,\bm{q}}^{\mathrm{V}} +A_{n,\bm{q}}^{\mathrm{H}}$.


\subsubsection{Pathloss} The pathloss $L_{n,\bm{q}}$ between BS $n$ and the user location $\bm{q}$ is a function of their distance and given by:
\begin{equation} \label{eqn:Pathloss}
L_{n,\bm{q}} = a_{\bm{q}} + b_{\bm{q}} \log_{10}\left[\| \bm{q} - \bm{p}_n \|^2 + (h_{\bm{q}} -h_{n,\mathrm{B}})^2 \right]^{\frac{1}{2}},
\end{equation}
where $a_{\bm{q}}$ depends on the carrier frequency and $b_{\bm{q}}$ relates to the line-of-sight condition and the pathloss exponent, which depends on the BS deployment feature and the user height at $\bm{q}$. In our case study in Section \ref{case-study}, we utilize practical values for the constants $a_{\bm{q}}$ and $b_{\bm{q}}$ that are adopted from the 3GPP studies \cite{3GPP36777,3GPP38901}.

In the remainder of the manuscript, we assume that the BS location $\bm{p}_n$ and the azimuth orientation $\phi_n$ are fixed for all $n \in \{1, \cdots, N\}$. We optimize over the vertical antenna tilts, cell partitioning, and BS transmission powers for multiple performance metrics that are introduced in the next two sections.
\section{Optimal RSS in Cellular Networks}\label{RSS-Chapter}

\subsection{Problem Formulation}\label{RSS-Problem-Formulation}

Our aim in this section is to optimize the average RSS across all users within the target region $Q$. The RSS from BS $n$, measured in dBm, provided at the user location $\bm{q}$ is given by:
\begin{multline}\label{RSS-dBm}
\mathtt{RSS_{dBm}^{(n)}}(\bm{q}; \theta_n)= \rho_n + A_{n,\bm{q}} - L_{n,\bm{q}}
 = \rho_n + A_{\textrm{max}} \\ - \frac{12}{\theta^2_{\text{3dB}}} \left[ \theta_{n,\bm{q}} - \theta_n \right]^2 - \frac{12}{\phi^2_{\text{3dB}}} \left[ \phi_{n,\bm{q}} - \phi_n \right]^2 
  - a_{\bm{q}} \\ - b_{\bm{q}} \log_{10}\left[\| \bm{q} - \bm{p}_n \|^2 + (h_{\bm{q}} -h_{n,\mathrm{B}})^2 \right]^{\frac{1}{2}}.
\end{multline}
The overall performance function, i.e., the RSS averaged over all network users, is given by:
\begin{equation}\label{RSS-objective-function}
    \Phi_{\mathtt{RSS}}(\bm{V}, \bm{\Theta}) = \sum_{n=1}^{N} \int_{V_n} \mathtt{RSS_{dBm}^{(n)}}(\bm{q}; \bm{\Theta}) \lambda(\bm{q}) d\bm{q}.
\end{equation}
In what follows, we seek to maximize the performance function $\Phi_{\mathtt{RSS}}$ in Eq. (\ref{RSS-objective-function}) over the cell partitioning $\bm{V}$ and BS vertical antenna tilts $\bm{\Theta}$.

\begin{Remark}
Due to the absence of interference from neighboring cells, optimizing $\Phi_{\mathtt{RSS}}$ w.r.t. the BS transmission powers $\bm{\rho}$ always reduces to allocating the maximum transmission power $\rho_{\max}$ to each BS. Hence, we only optimize over the cell partitioning and BS vertical antenna tilts while assuming that the BS transmission powers $\bm{\rho}$ are given and fixed.
\end{Remark}

\begin{Remark}
The RSS function in Eq. (\ref{RSS-dBm}) is not necessarily a non-increasing function of the distance between the BS and the user. This is because while moving away from the BS worsens the pathloss component, it may lead to a better antenna gain and thus, an overall RSS value.
\end{Remark}

\subsection{Analytical Framework}\label{RSS-Analytical-Framework}

Our goal is to optimize the performance function $\Phi_{\mathtt{RSS}}$ over variables $\bm{V}$ and $\bm{\Theta}$. Not only does the optimal choice of each variable depend on the value of the other, but also this optimization problem is NP-hard. Our approach is to design an alternating optimization algorithm that iterates between updating $\bm{V}$ and $\bm{\Theta}$. 
In quantization theory, variations of the Lloyd algorithm \cite{lloyd1982least, gray1998quantization} have been used to solve similar optimization problems. Inspired by quantization theory, we need to: (i) find the optimal cell partitioning $\bm{V}$ given a set of BS vertical antenna tilts $\bm{\Theta}$; and (ii) find the optimal vertical antenna tilts $\bm{\Theta}$ for a given cell partitioning $\bm{V}$. The solution of the first task is a generalized Voronoi tessellation \cite{boots2009spatial, du1999centroidal} carried out via the following proposition:
\begin{Proposition}\label{optimal-V}
For a given set of BS vertical antenna tilts $\bm{\Theta}$, the optimal cell partitioning $\bm{V}^*(\bm{\Theta}) = \big(V_1^*(\bm{\Theta}), \cdots, V_N^*(\bm{\Theta})\big)$ that maximizes the performance function $\Phi_{\mathtt{RSS}}$ is given by:
\begin{multline}\label{optimal-cell-partitioning}
    V_n^*(\bm{\Theta}) = \big\{\bm{q} \in Q \mid \mathtt{RSS_{dBm}^{(n)}}(\bm{q}; \theta_n) \geq \mathtt{RSS_{dBm}^{(k)}}(\bm{q}; \theta_k), \\ \textrm{ for all } 1 \leq k \leq N \big\},
\end{multline}
for each $n \in \{1, \cdots, N\}$. The ties can be broken arbitrarily.
\end{Proposition}
\textit{Proof. }Let $\bm{W} = (W_1, \cdots, W_N)$ be any arbitrary cell partitioning of the target region $Q$. Then: 
\begin{align*}
    \Phi_{\mathtt{dBm}}(\bm{W},\bm{\Theta}) &= \sum_{n=1}^{N} \int_{W_n} \mathtt{RSS_{dBm}^{(n)}} (\bm{q}; \bm{\Theta}) \lambda(\bm{q}) d\bm{q} \\& \leq \sum_{n=1}^{N} \int_{W_n} \max_k \Big[\mathtt{RSS_{dBm}^{(k)}} (\bm{q}; \bm{\Theta})\Big] \lambda(\bm{q}) d\bm{q} \\
    &=\int_Q \max_k \Big[\mathtt{RSS_{dBm}^{(k)}} (\bm{q}; \bm{\Theta})\Big] \lambda(\bm{q}) d\bm{q} \\&=\sum_{n=1}^{N} \int_{V_n^*} \max_k \Big[\mathtt{RSS_{dBm}^{(k)}} (\bm{q}; \bm{\Theta}))\Big] \lambda(\bm{q}) d\bm{q} \\
    &=\sum_{n=1}^{N} \int_{V_n^*} \mathtt{RSS_{dBm}^{(n)}} (\bm{q}; \bm{\Theta}) \lambda(\bm{q}) d\bm{q} \\&= \Phi_{\mathtt{dBm}}(\bm{V}^*, \bm{\Theta}),
\end{align*}
i.e., $\bm{V}^*$ achieves a performance no less than any other partitioning $\bm{W}$ and is optimal.$\hfill\blacksquare$

For the second task, our approach is to apply gradient ascent to find the optimal BS vertical antenna tilts $\bm{\Theta}$ for a given cell partitioning $\bm{V}$. Gradient ascent is a first-order optimization algorithm that iteratively refines the estimate of a locally optimal $\bm{\Theta}$ by following the direction of the gradient.
\begin{Proposition}\label{rss-grad-eq}
The partial derivative of the performance function $\Phi_{\mathtt{RSS}}$ w.r.t. $\theta_n$ is given by:
\begin{align}\label{eqn:derivativePhidBm}
\frac{\partial \Phi(\bm{V},\bm{\Theta})}{\partial \theta_n}  =   \frac{24}{\theta^2_{\text{3dB}}} \Bigg\{ \sum_{u=1}^{N_U}
      \int_{V_n(\mathbf{\Theta})\cap Q_u} \!\!\! (\theta_{n,\bm{q}}-\theta_n) \lambda(\bm{q}) d\bm{q}   \nonumber\\+ \int_{V_n(\mathbf{\Theta})\cap Q_G} \!\!\! (\theta_{n,\bm{q}}-\theta_n) \lambda(\bm{q}) d\bm{q} \Bigg\}.
\end{align}
\end{Proposition}
\textit{Proof. }The partial derivative of Eq. (\ref{RSS-objective-function}) w.r.t. $\theta_n$ consists of two components: (i) the derivative of the integrand; and (ii) the integral over the boundaries of $V_n$ and its neighboring regions. For any point $\bm{q}$ on the boundary of neighboring regions $V_n$ and $V_m$, the normal outward vectors have opposite directions and we have $\mathtt{RSS_{dBm}^{(n)}}(\bm{q}; \bm{\Theta}) = \mathtt{RSS_{dBm}^{(m)}}(\bm{q}; \bm{\Theta})$; thus, the sum of elements in the second component is zero \cite{GuoJaf2016}. The first component evaluates to:
\begin{multline}\label{eqn:derivativePhi}
    \frac{\partial \Phi(\mathbf{V},\mathbf{\Theta})}{\partial \theta_n} = \int_{V_n(\mathbf{\Theta})} \frac{\partial}{\partial \theta_n} \mathtt{RSS_{dBm}^{(n)}}(\bm{q}; \theta_n) \lambda(\bm{q})d\bm{q}  \\
    \stackrel{(\text{a})}{=}  \frac{24}{\theta^2_{\text{3dB}}} \Bigg\{ \sum_{u=1}^{N_U}
      \int_{V_n(\mathbf{\Theta})\cap Q_u}  (\theta_{n,\bm{q}}-\theta_n) \lambda(\bm{q}) d\bm{q}   
      \\+  \int_{V_n(\mathbf{\Theta})\cap Q_G}  (\theta_{n,\bm{q}}-\theta_n) \lambda(\bm{q}) d\bm{q} \Bigg\},
\end{multline}
where (a) follows from the definition of $Q = Q_U \bigcup Q_G$, and the proof is complete. $\hfill\blacksquare$

\subsection{Proposed Algorithm}\label{RSS-Algorithm}

With our two tasks accomplished in Propositions \ref{optimal-V} and \ref{rss-grad-eq}, we propose the maximum-RSS vertical antenna tilt (Max-RSS-VAT) iterative optimization algorithm outlined in Algorithm \ref{BS_VAT_Algorithm}.

\begin{algorithm}[ht!]
\SetAlgoLined
\SetKwRepeat{Do}{do}{while}
\KwResult{Optimal BS vertical antenna tilts $\bm{\Theta}^*$ and cell partitioning $\bm{V}^*$.}
\textbf{Input:} Initial BS vertical antenna tilts $\mathbf{\Theta}$ and cell partitioning $\bm{V}$, learning rate $\eta_0\in (0,1)$, convergence error thresholds $\epsilon_1, \epsilon_2\in \mathbb{R}^+$, constant $\kappa \in (0, 1)$\;

\Do{$\frac{\Phi_{\mathtt{RSS}}^{\textrm{(new)}} - \Phi_{\mathtt{RSS}}^{\textrm{(old)}}}{\Phi_{\mathtt{RSS}}^{\textrm{(old)}}} \geq \epsilon_2$}
{
-- Calculate  $\Phi_{\mathtt{RSS}}^{\textrm{(old)}} = \Phi_{\mathtt{RSS}}\left(\bm{V},\mathbf{\Theta}\right)$\;
-- Update the cell $V_n$ according to Eq. (\ref{optimal-cell-partitioning}) for each $n \in \{1, \cdots, N\}$\;
-- Set $\eta \gets \eta_0$\;
\Do{$\frac{\Phi_{\textrm{e}} - \Phi_{\textrm{s}}}{\Phi_{\textrm{s}}} \geq \epsilon_1$}
{
-- Calculate  $\Phi_{\textrm{s}} = \Phi_{\mathtt{RSS}}\left(\bm{V},\mathbf{\Theta}\right)$\;
-- Calculate $\frac{\partial \Phi_{\mathtt{RSS}}(\mathbf{V},\mathbf{\Theta})}{\partial \theta_n}$ according to Eq. (\ref{eqn:derivativePhidBm}) for each $n \in \{1, \cdots, N\}$\;
-- $\eta \gets \eta \times \kappa$\;
-- $\mathbf{\Theta} \gets \mathbf{\Theta} + \eta \nabla_{\mathbf{\Theta}} \Phi_{\mathtt{RSS}}(\bm{V},\mathbf{\Theta})$\;
-- Calculate $\Phi_{\textrm{e}} = \Phi_{\mathtt{RSS}}\left(\bm{V},\mathbf{\Theta}\right)$\;
}
-- Calculate  $\Phi_{\mathtt{RSS}}^{\textrm{(new)}} = \Phi_{\mathtt{RSS}}\left(\bm{V},\mathbf{\Theta}\right)$\;
}
 \caption{Maximum-RSS vertical antenna tilt (Max-RSS-VAT) optimization}
 \label{BS_VAT_Algorithm}
\end{algorithm}

\begin{Proposition}\label{BS-VAT-convergence}
The Max-RSS-VAT algorithm is an iterative improvement algorithm and converges.
\end{Proposition}
\textit{Proof. }Proposition \ref{optimal-V} indicates that updating the cell $V_n$ according to Eq. (\ref{optimal-cell-partitioning}), as it is done in the Max-RSS-VAT algorithm, yields the optimal cell partitioning for a given value of $\bm{\Theta}$; thus, the performance function $\Phi_{\mathtt{RSS}}$ will not decrease as a result of this update rule. The Max-RSS-VAT algorithm updates the vertical antenna tilts $\bm{\Theta}$ using gradient ascent where the learning rate at time $t$ is given by $\eta_t = \eta_0 \times \kappa^t$. Because $\sum_{t=1}^{\infty}\eta^2_t < \sum_{t=1}^{\infty}\eta_t = \frac{\kappa}{1 - \kappa}\eta_0 < \infty$, the gradient ascent is guaranteed to converge \cite{goodfellow2016deep} and does not decrease the performance function $\Phi_{\mathtt{RSS}}$. Hence, the Max-RSS-VAT algorithm generates a sequence of non-decreasing performance function values that are also upper bounded because of the limited transmission power at each BS; thus, the algorithm converges.  $\hfill\blacksquare$
\section{Optimal SINR in Cellular Networks}\label{SINR-Chapter}

\subsection{Problem Formulation}\label{SINR-Problem-Formulation}

Our goal in this section is to optimize the average signal-to-interference-plus-noise ratio (SINR) across all users within the target region $Q$. Not only is this optimization performed over the cell partitioning $\bm{V}$ and BS vertical antenna tilts $\bm{\Theta}$, but also this is done over BS transmission power values $\bm{\rho}$. Indeed, unlike the case of RSS in Section \ref{RSS-Chapter}, BS transmission power values play a crucial role because of the interference from neighboring cells. Using the definition of $\mathtt{RSS_{dBm}^{(n)}}$ in Eq. (\ref{RSS-dBm}), we define:
\begin{equation}\label{eqn:SINR_dB}
\mathtt{SINR^{(n)}_{dB}} (\bm{q}; \mathbf{\Theta}, \bm{\rho}) = 10 \log_{10} \frac{10^{\frac{1}{10}\mathtt{RSS^{(n)}_{dBm}} (\bm{q}; \theta_n, \rho_n)}}{\sum_{j\neq n}^{} 10^{\frac{1}{10}\mathtt{RSS^{(j)}_{dBm}} (\bm{q}; \theta_j, \rho_j)} + \sigma^2}  
\end{equation}
where $\sigma^2$ denotes the noise variance in linear units. The performance function, which is the SINR measured in dB and averaged over all network users, is given by:
\begin{align}\label{SINR-objective}
    \Phi_{\mathtt{SINR}}(\bm{V}, \mathbf{\Theta}, \bm{\rho}) &=  \sum_{n=1}^{N} \int_{V_n} \mathtt{SINR_{dB}^{(n)}}(\bm{q}; \mathbf{\Theta}, \bm{\rho}) \lambda(\bm{q}) d\bm{q}, \\
    \textrm{s.t. } \rho_n &\leq \rho_{\max} \qquad \forall n\in\{1,\cdots, N\}, \label{SINR-objective-constraint}
\end{align}
where the constraint in Eq. (\ref{SINR-objective-constraint}) comes from the fact that for any BS, say $n$, the transmission power $\rho_n$ measured in dBm cannot exceed $\rho_{\max}$. In what follows, we aim to optimize the performance function $\Phi_{\mathtt{SINR}}$ over the cell partitioning, BS vertical antenna tilts, and BS transmission powers.

\subsection{Analytical Framework}\label{SINR-Analytical-Framework}

Our approach to optimize the performance function $\Phi_{\mathtt{SINR}}$ over variables $\bm{V}$, $\bm{\Theta}$, and $\bm{\rho}$ is via an alternating optimization algorithm that iteratively optimizes each variable while the other two are held fixed. This goal is carried out over the following three steps: (i) find the optimal cell partitioning $\bm{V}$ for a given BS vertical antenna tilt $\bm{\Theta}$ and transmission power $\bm{\rho}$; (ii) find the optimal antenna tilts $\bm{\Theta}$ for a given cell partitioning and BS transmission power $\bm{\rho}$; and (iii) find the optimal BS transmission power $\bm{\rho}$ for a given cell partitioning $\bm{V}$ and vertical antenna tilts $\bm{\Theta}$. The first step is accomplished in the following proposition. 
\begin{Proposition}\label{optimal-V-SINR}
For a given set of BS vertical antenna tilts $\bm{\Theta}$ and transmission power values $\bm{\rho}$, the optimal cell partitioning $\bm{V}^*(\bm{\Theta}, \bm{\rho}) = \big(V_1^*(\bm{\Theta}, \bm{\rho}), \cdots, V_N^*(\bm{\Theta}, \bm{\rho})\big)$ that maximizes the performance function $\Phi_{\mathtt{SINR}}$ is given by:
\begin{multline}\label{optimal-cell-partitioning-SINR}
    \!\!\! V_n^*(\bm{\Theta}, \bm{\rho}) \!=\! \big\{\bm{q} \in Q \mid \mathtt{RSS_{dBm}^{(n)}}(\bm{q}; \theta_n, \rho_n) \geq \mathtt{RSS_{dBm}^{(k)}}(\bm{q}; \theta_k, \rho_k), \\ \textrm{ for all } 1 \leq k \leq N \big\},
\end{multline}
for each $n \in \{1, \cdots, N\}$. The ties can be broken arbitrarily.
\end{Proposition}
\noindent The proof of Proposition \ref{optimal-V-SINR} is provided in Appendix \ref{Appendix_A}.

For the second step, we aim to apply gradient ascent to find the optimal $\bm{\Theta}$ for a given cell partitioning and BS transmission power. The following proposition provides the main ingredient needed for this process.
\begin{Proposition}\label{gradient-equation-SINR}
The derivative of Eq. (\ref{SINR-objective}) w.r.t. the BS vertical antenna tilt $\theta_n$ is given by:
\begin{multline}\label{gradient-equation-SINR-equation}
     \frac{\partial \Phi_{\mathtt{SINR}}(\bm{V}, \bm{\Theta}, \bm{\rho})}{\partial \theta_n} =  \frac{24}{\theta^2_{\mathrm{3dB}}} \Bigg\{ \sum_{u=1}^{N_U} \int_{V_n(\mathbf{\Theta}, \bm{\rho})\cap Q_u} \!\!\!\!\!\!\!\!\!\!\!\! (\theta_{n,\bm{q}}-\theta_n) \lambda(\bm{q}) d\bm{q}  
    \\ +  \int_{V_n(\mathbf{\Theta}, \bm{\rho})\cap Q_G} \!\!\!\! (\theta_{n,\bm{q}}-\theta_n) \lambda(\bm{q}) d\bm{q} \Bigg\}  -\frac{24}{\theta^2_{\mathrm{3dB}}} \sum_{i\neq n}^{}  \Bigg\{ \\ \sum_{u=1}^{N_U} \int_{V_i(\mathbf{\Theta}, \bm{\rho})\cap Q_u}  \frac{(\theta_{n,\bm{q}}-\theta_n) \cdot 10^{\frac{1}{10}\mathtt{RSS_{dBm}^{(n)}} (\bm{q}; \theta_n, \rho_n)}}{{\sum_{j\neq i}^{} 10^{\frac{1}{10}\mathtt{RSS_{dBm}^{(j)}} (\bm{q}; \theta_j, \rho_j)} + \sigma^2}} \lambda(\bm{q}) d\bm{q} 
    \\+ \int_{V_i(\mathbf{\Theta}, \bm{\rho})\cap Q_G}  \frac{(\theta_{n,\bm{q}}-\theta_n) \cdot 10^{\frac{1}{10}\mathtt{RSS_{dBm}^{(n)}} (\bm{q}; \theta_n, \rho_n)}}{{\sum_{j\neq i}^{} 10^{\frac{1}{10}\mathtt{RSS_{dBm}^{(j)}} (\bm{q}; \theta_j,\rho_j)} + \sigma^2}} \lambda(\bm{q}) d\bm{q} \Bigg\}.
\end{multline}
\end{Proposition}
\textit{Proof. }Similar to the proof of Proposition \ref{rss-grad-eq}, it can be shown that the partial derivative in Eq. (\ref{gradient-equation-SINR-equation}) has two components and the second component is zero. This is because, according to Eq. (\ref{optimal-cell-partitioning-SINR-proof}) in Appendix \ref{Appendix_A}, for any point $\bm{q}$ on the boundary of neighboring regions $V_n$ and $V_m$, we have $\mathtt{SINR_{dBm}^{(n)}}(\bm{q}; \theta_n, \rho_n) = \mathtt{SINR_{dBm}^{(m)}}(\bm{q}; \theta_m, \rho_m)$ and the unit outward normal vectors have opposite directions \cite{GuoJaf2016}. Thus:
\begin{multline}\label{eqn:derivativePhi-SINR}
    \frac{\partial \Phi_{\mathtt{SINR}}(\bm{V},\mathbf{\Theta}, \bm{\rho})}{\partial \theta_n} = 
    \sum_{i=1}^{N} \int_{V_i(\mathbf{\Theta}, \bm{\rho})} \!\!\!\frac{\partial \mathtt{SINR_{dB}^{(i)}} (\bm{q}; \mathbf{\Theta}, \bm{\rho})}{\partial \theta_n}  \lambda(\bm{q})d\bm{q} \\
     =
    \int_{V_n(\mathbf{\Theta}, \bm{\rho})} \frac{\partial}{\partial \theta_n} \mathtt{SINR_{dB}^{(n)}} (\bm{q}; \mathbf{\Theta}, \bm{\rho}) \lambda(\bm{q})d\bm{q} 
      \\ + \sum_{i\neq n}^{} \int_{V_i(\mathbf{\Theta}, \bm{\rho})} \frac{\partial}{\partial \theta_n} \mathtt{SINR_{dB}^{(i)}} (\bm{q}; \mathbf{\Theta}, \bm{\rho}) \lambda(\bm{q})d\bm{q}.
\end{multline}
Eq. (\ref{gradient-equation-SINR-equation}) is then derived via straightforward algebraic operations on Eq. (\ref{eqn:derivativePhi-SINR}) and using the definition of SINR in Eq. (\ref{eqn:SINR_dB}), which concludes the proof. $\hfill\blacksquare$

Finally, for the third step, we optimize the BS transmission power $\bm{\rho}$ for a given cell partitioning $\bm{V}$ and vertical antenna tilts $\bm{\Theta}$. We utilize the gradient projection method, a variation of the gradient ascent algorithm that keeps the power of each BS lower than its maximum possible power. To this end, we require the gradient formula given below.
\begin{Proposition}\label{sinr-power-allocation-gradient}
The derivative of Eq. (\ref{SINR-objective}) w.r.t. the BS transmission power $\rho_n$ is given by:
\begin{multline}\label{sinr-power-allocation-gradient-equation}
     \frac{\partial \Phi_{\mathtt{SINR}}(\bm{V}, \bm{\Theta}, \bm{\rho})}{\partial \rho_n} =   \Bigg\{ \sum_{u=1}^{N_U} \int_{V_n(\mathbf{\Theta}, \bm{\rho})\cap Q_u}   \lambda(\bm{q}) d\bm{q}  
    \\+  \int_{V_n(\mathbf{\Theta}, \bm{\rho})\cap Q_G}  \lambda(\bm{q}) d\bm{q} \Bigg\}   - \sum_{i\neq n}^{}  \Bigg\{ \\ \sum_{u=1}^{N_U} \int_{V_i(\mathbf{\Theta}, \bm{\rho})\cap Q_u} \!\!\!\!\!\!\!\!\! \!\!\!\!\frac{\mathtt{RSS_{lin}^{(n)}}(\bm{q};\theta_n,\rho_n) \times \mathtt{SINR_{lin}^{(i)}}(\bm{q}; \bm{\Theta},\bm{\rho})}{\mathtt{RSS_{lin}^{(i)}}(\bm{q};\theta_i,\rho_i)} \lambda(\bm{q}) d\bm{q}
    \\+ \int_{V_i(\mathbf{\Theta}, \bm{\rho})\cap Q_G} \!\!\!\!\!\!\!\!\!\!\!\!\! \frac{\mathtt{RSS_{lin}^{(n)}}(\bm{q};\theta_n,\rho_n) \times \mathtt{SINR_{lin}^{(i)}}(\bm{q}; \bm{\Theta},\bm{\rho})}{\mathtt{RSS_{lin}^{(i)}}(\bm{q};\theta_i,\rho_i)} \lambda(\bm{q}) d\bm{q} \Bigg\}.
\end{multline}
\end{Proposition}
\noindent The proof of Proposition \ref{sinr-power-allocation-gradient} is provided in Appendix \ref{Appendix_B}.

In the remainder of this section, we embed Propositions \ref{optimal-V-SINR}, \ref{gradient-equation-SINR}, and \ref{sinr-power-allocation-gradient} into an alternating optimization algorithm that maximizes the average SINR across all network users.

\subsection{Proposed Algorithm}\label{SINR-Algorithm}

Propositions \ref{optimal-V-SINR}, \ref{gradient-equation-SINR}, and \ref{sinr-power-allocation-gradient} provide the main ingredients required for the three-step maximum-SINR power allocation and vertical antenna tilt (Max-SINR-PA-VAT)
optimization process presented in Algorithm \ref{BS_PA_VAT_Algorithm}. While BS vertical antenna tilts $\bm{\Theta}$ are optimized via gradient ascent, as shown in Algorithm \ref{BS_PA_VAT_Algorithm}, the BS transmission powers $\bm{\rho}$ are optimized via the gradient projection method with the projection operator $P_{\bm{\Lambda}}(.)$ that projects the updated $\bm{\rho}$ onto the subspace $\bm{\Lambda} = (-\infty, \rho_{\max}]^N$. This is done to make sure that the range of all transmission power values remain in the feasible set and satisfy the constraint in Eq. (\ref{SINR-objective-constraint}).

\begin{algorithm}[ht!]
\SetAlgoLined
\SetKwRepeat{Do}{do}{while}
\KwResult{Optimal cell partitioning $\bm{V}^*$, BS antenna tilts $\bm{\Theta}^*$ and transmission power $\bm{\rho}^*$.}

\textbf{Input:} Initial cell partitioning $\bm{V}$, BS vertical antenna tilts $\mathbf{\Theta}$ and transmission power $\bm{\rho}$, 
maximum BS transmission power $\rho_{\max}$, learning rates $\eta_0, \eta'_0 \in (0,1)$, 
convergence error thresholds $\epsilon_1, \epsilon_2, \epsilon_3 \in \mathbb{R}^+$, constant $\kappa \in  (0, 1)$\;

\Do{$\frac{\Phi_{\mathtt{SINR}}^{\textrm{(new)}} - \Phi_{\mathtt{SINR}}^{\textrm{(old)}}}{\Phi_{\mathtt{SINR}}^{\textrm{(old)}}} \geq \epsilon_3$}
{
-- Calculate  $\Phi_{\mathtt{SINR}}^{\textrm{(old)}} = \Phi_{\mathtt{SINR}}\left(\bm{V},\mathbf{\Theta}, \bm{\rho}\right)$\;
-- Update the cell $V_n$ according to Eq. (\ref{optimal-cell-partitioning-SINR}) for each $n \in \{1, \cdots, N\}$\;
-- Set $\eta \gets \eta_0$\;
\Do{$\frac{\Phi_{\textrm{e}} - \Phi_{\textrm{s}}}{\Phi_{\textrm{s}}} \geq \epsilon_1$}
{
-- Calculate  $\Phi_{\textrm{s}} = \Phi_{\mathtt{SINR}}\left(\bm{V},\mathbf{\Theta},\bm{\rho}\right)$\;
-- Calculate $\frac{\partial \Phi_{\mathtt{SINR}}(\mathbf{V},\mathbf{\Theta},\bm{\rho})}{\partial \theta_n}$ according to Eq. (\ref{gradient-equation-SINR-equation}) for each $n \in \{1, \cdots, N\}$\;
-- $\eta \gets \eta \times \kappa$\;
-- $\mathbf{\Theta} \gets \mathbf{\Theta} + \eta \nabla_{\mathbf{\Theta}} \Phi_{\mathtt{SINR}}(\bm{V},\mathbf{\Theta},\bm{\rho})$\;
-- Calculate $\Phi_{\textrm{e}} = \Phi_{\mathtt{SINR}}\left(\bm{V},\mathbf{\Theta}, \bm{\rho}\right)$\;
}
-- Set $\eta \gets \eta'_0$\;
\Do{$\frac{\Phi_{\textrm{e}} - \Phi_{\textrm{s}}}{\Phi_{\textrm{s}}} \geq \epsilon_2$}
{
-- Calculate  $\Phi_{\textrm{s}} = \Phi_{\mathtt{SINR}}\left(\bm{V},\mathbf{\Theta}, \bm{\rho}\right)$\;
-- Calculate $\frac{\partial \Phi_{\mathtt{SINR}}(\mathbf{V},\mathbf{\Theta}, \bm{\rho})}{\partial \rho_n}$ according to Eq. (\ref{sinr-power-allocation-gradient-equation}) for each $n \in \{1, \cdots, N\}$\;
-- $\eta \gets \eta \times \kappa$\;
-- $\bm{\rho} \gets P_{\bm{\Lambda}}\big(\bm{\rho} + \eta \nabla_{\bm{\rho}} \Phi_{\mathtt{SINR}}(\bm{V},\mathbf{\Theta}, \bm{\rho})\big)$\;
-- Calculate $\Phi_{\textrm{e}} = \Phi_{\mathtt{SINR}}\left(\bm{V},\mathbf{\Theta}, \bm{\rho}\right)$\;
}
-- Calculate  $\Phi_{\mathtt{SINR}}^{\textrm{(new)}} = \Phi_{\mathtt{SINR}}\left(\bm{V},\mathbf{\Theta}, \bm{\rho}\right)$\;
}
 \caption{Maximum-SINR power allocation and vertical antenna tilt (Max-SINR-PA-VAT) optimization}
 \label{BS_PA_VAT_Algorithm}
\end{algorithm}

\begin{Proposition}\label{BS-PA-VAT-convergence}
    The Max-SINR-PA-VAT algorithm is an iterative improvement algorithm and converges.
\end{Proposition}

{\it Proof. } The Max-SINR-PA-VAT algorithm iteratively updates the parameters $\bm{V}$, $\bm{\Theta}$, and $\bm{\rho}$. Updating the cell partitioning $\bm{V}$ according to Eq. (\ref{optimal-cell-partitioning-SINR}) does not decrease the performance function $\Phi_{\mathtt{SINR}}$ because Proposition \ref{optimal-V-SINR} guarantees its optimality for a given $\bm{\Theta}$ and $\bm{\rho}$. A similar argument to the one presented in Proposition \ref{BS-VAT-convergence} suggests that updating $\bm{\Theta}$ and $\bm{\rho}$ using the gradient ascent and the gradient projection methods, respectively, will not result in a decrease in the performance function. This indicates that Algorithm \ref{BS_PA_VAT_Algorithm} produces a sequence of performance function values that are non-decreasing and upper-bounded, as a result of the finite transmission power at each BS; thus, it converges. $\hfill\blacksquare$

The above general framework, inspired by quantization theory, works for any performance function for which the required gradients can be calculated. To demonstrate the general capability of our proposed solution, in the sequel, we introduce and optimize two generalized performance functions: the max-product SINR and the soft max-min SINR \cite{nikbakht2020unsupervised}. These performance functions avoid the occasional disparities among individual network users that can happen when optimizing the performance function $\Phi_{\mathtt{SINR}}$.




\subsection{Generalization to Max-Product SINR}\label{MP-Subsection}

\subsubsection{Performance Function}\label{MP-Objective}
The goal of the max-product performance function is to maximize the product of the SINRs. 
The max-product proxy performance function is defined as:
\begin{align}\label{max-product-objective}
    \Phi_{{\sf MP}}(\bm{V}, \mathbf{\Theta}, \bm{\rho}) &=  \sum_{n=1}^{N} \int_{V_n} \gamma^{(n)}_{{\sf MP}} (\bm{q}; \mathbf{\Theta}, \bm{\rho}) \lambda(\bm{q}) d\bm{q}, \\
    \textrm{s.t. } \rho_n &\leq \rho_{\max} \qquad \forall n \in \{1,\cdots, N\}, \label{max-product-objective-constraint}
\end{align}
where
\begin{align}\label{Max-product}
\gamma_{\sf MP}^{(n)}(\bm{q}; \mathbf{\Theta}, \bm{\rho}) = - \log \! \left[ \mu + \frac{1}{(\mathtt{SINR^{(n)}_{lin}}(\bm{q}; \mathbf{\Theta}, \bm{\rho}) + \nu )} \right].
\end{align}
The offset $\nu$ prevents the performance from being dominated by users with very low SINRs. The offset $\mu$ plays a similar role for high SINRs. 
Note that for the special case of $\mu = \nu = 0$, $\Phi_{{\sf MP}}$ in \eqref{max-product-objective} boils down to $\Phi_{\mathtt{SINR}}$ in \eqref{SINR-objective} except for a constant multiplier. As a result, 
\eqref{max-product-objective} can be considered as a generalization of \eqref{SINR-objective}.

\subsubsection{Optimal Configuration}\label{MP-Optimal-Configuration}

The iterative process for maximizing the constrained performance function described in Eqs. (\ref{max-product-objective}) and (\ref{max-product-objective-constraint}) over variables $\bm{V}$, $\bm{\Theta}$, and $\bm{\rho}$ is similar to the one outlined in Section \ref{SINR-Algorithm}. This process requires determining the optimality conditions for each variable while keeping the other two variables constant.

\begin{Proposition}\label{MP-optimal-cell}
For a given BS vertical antenna tilts $\bm{\Theta}$ and transmission powers $\bm{\rho}$, the optimal cell partitioning $\bm{V}^*(\bm{\Theta}, \bm{\rho}) = \left(V^*_1(\bm{\Theta}, \bm{\rho}), \cdots, V_N^*(\bm{\Theta}, \bm{\rho})\right)$ that maximizes $\Phi_{{\sf MP}}$ is given by:
\begin{multline}\label{optimal-cell-partitioning-MP}
    \!\!\!\!\! V_n^*(\bm{\Theta}, \bm{\rho}) = \big\{\bm{q} \in Q \mid \mathtt{RSS_{dBm}^{(n)}}(\bm{q}; \theta_n, \rho_n) \geq \mathtt{RSS_{dBm}^{(k)}}(\bm{q}; \theta_k, \rho_k), \\ \textrm{ for all } 1 \leq k \leq N \big\}.,
\end{multline}        
for each $n \in \{1, \cdots, N\}$. The ties can be broken arbitrarily.
\end{Proposition}
\textit{Proof. }
Since $\mu$ and $\nu$ are constants, $\gamma_{\sf MP}^{(n)}(\bm{q}; \mathbf{\Theta}, \bm{\rho}) \geq \gamma_{\sf MP}^{(k)}(\bm{q}; \mathbf{\Theta}, \bm{\rho})$ for all $k \neq n$ is the same as Eq. \eqref{optimal-cell-partitioning-SINR-proof} in Appendix \ref{Appendix_A}; therefore, the rest of the proof follows from that of Proposition \ref{optimal-V-SINR}. $\hfill\blacksquare$ 

Next, we provide the partial derivative expression for $\Phi_{{\sf MP}}$ w.r.t. the BS $n$'s antenna tilt $\theta_n$.

\begin{Proposition}\label{gradient-theta-MP}
    The partial derivative of the performance function $\Phi_{{\sf MP}}$ w.r.t. the BS $n$'s vertical antenna tilt $\theta_n$ is given by Eq. (\ref{gradient-theta-MP-equation}), on top of the next page, 
\begin{figure*}[t!]
\begin{multline}\label{gradient-theta-MP-equation}
    \frac{\partial \Phi_{{\sf MP}}}{\partial \theta_n} =   \int_{V_n}  \frac{ \SINR^{(n)}_{\mathtt{lin}} \cdot \frac{2.4 \log 10}{\theta^2_{\textrm{3dB}}} \cdot \left(\theta_{n,\bm{q}} - \theta_n \right)}{\left[\SINR^{(n)}_{\mathtt{lin}} + \nu \right] \cdot \left[ 1 + \mu \left( \SINR^{(n)}_{\mathtt{lin}} + \nu \right) \right] }  \lambda(\bm{q})d\bm{q} \\
    - \sum_{i\neq n}   \int_{V_i }  \frac{\SINR^{(i)}_{\mathtt{lin}} \cdot \frac{2.4 \log 10}{\theta^2_{\text{3dB}}} \cdot \left(\theta_{n,q} - \theta_n \right)\cdot \RSS^{(n)}_{\mathtt{lin}}  }{\left[\SINR^{(i)}_{\mathtt{lin}} + \nu \right] \cdot 
    \left[ 1 + \mu \left( \SINR^{(i)}_{\mathtt{lin}} + \nu \right) \right] \cdot \left[ {\sum_{j\neq i}^{} \RSS^{(j)}_{\mathtt{lin}}  + \sigma^2_{\mathtt{lin}}} \right] }   \lambda(\bm{q})d\bm{q}.  
\end{multline}   
\end{figure*}
where for the sake of brevity of the notation, the dependence of the variables $\mathtt{SINR}_{\mathtt{lin}}^{(n)}$, $\mathtt{RSS}_{\mathtt{lin}}^{(n)}$, and $V_n$ on $\bm{\Theta}$ and $\bm{\rho}$ is omitted.
\end{Proposition}
\noindent The proof is similar to that of Proposition \ref{gradient-equation-SINR} and is omitted.

Next, we provide the partial derivative of $\Phi_{{\sf MP}}$ w.r.t. $\rho_n$.

\begin{Proposition}\label{gradient-formula-MP-power}
The partial derivative of $\Phi_{{\sf MP}}$ w.r.t. the BS transmission power $\rho_n$ is given by Eq. (\ref{gradient-MP-power}),  on top of the next page,
\begin{figure*}[t!]
    \begin{multline}\label{gradient-MP-power}
    \frac{\partial \Phi_{{\sf MP}}}{\partial \rho_n} =   \int_{V_n }  \frac{\SINR^{(n)}_{\mathtt{lin}} \cdot \frac{\log 10}{10}}{\left[\SINR^{(n)}_{\mathtt{lin}} + \nu \right] \cdot 
    \left[ 1 + \mu \left( \SINR^{(n)}_{\mathtt{lin}} + \nu \right) \right] }   \lambda(\bm{q})d\bm{q}   
    - \sum_{i\neq n}   \int_{V_i }     \frac{\left[\frac{1}{10}\log (10) \cdot \left[ \SINR^{(i)}_{\mathtt{lin}} \right]^2 \cdot \frac{\RSS^{(n)}_{\mathtt{lin}}}{\RSS^{(i)}_{\mathtt{lin}}}
    \right]}{\left[\SINR^{(i)}_{\mathtt{lin}} + \nu \right] \cdot 
    \left[ 1 + \mu \left( \SINR^{(i)}_{\mathtt{lin}} + \nu \right) \right] }  
     \lambda(\bm{q})d\bm{q}.
\end{multline}
\end{figure*}
where $\SINR^{(n)}_{\mathtt{lin}}(\bm{q}; \bm{\Theta}, \bm{\rho})$, $\RSS^{(n)}_{\mathtt{lin}}(\bm{q}; \theta_n, \rho_n)$, and $V_n(\bm{\Theta}, \bm{\rho})$ are abbreviated as
$\SINR^{(n)}_{\mathtt{lin}} $, $\RSS^{(n)}_{\mathtt{lin}}$, and $V_n$, respectively. 
\end{Proposition}
\noindent The proof is similar to that of Proposition \ref{sinr-power-allocation-gradient} and is omitted.

\subsubsection{Proposed Algorithm}\label{MP-Algorithm}

Using Propositions \ref{MP-optimal-cell}, \ref{gradient-theta-MP}, and \ref{gradient-formula-MP-power}, 
after a random initialization for the values of $\bm{V}$, $\bm{\Theta}$, and $\bm{\rho}$, our 
max-product power allocation and vertical antenna tilt (MP-PA-VAT) optimization algorithm, iterates over the following main three steps until its convergence criterion is met: 
\begin{itemize}
    \item Adjust the cell $V_n$ according to Eq. (\ref{optimal-cell-partitioning-MP}) for each $n \in \{1, \cdots, N\}$ while $\bm{\Theta}$ and $\bm{\rho}$ are fixed;
    \item Calculate the gradient $\nabla_{\bm{\Theta}}\Phi_{{\sf MP}}$ according to Proposition~\ref{gradient-theta-MP} and apply the gradient ascent algorithm to optimize $\bm{\Theta}$ while $\bm{V}$ and $\bm{\rho}$ are fixed;
    \item Compute the gradient vector $\nabla_{\bm{\rho}}\Phi_{{\sf MP}}$ according to Proposition~\ref{gradient-formula-MP-power} and use the projected gradient ascent algorithm to optimize $\bm{\rho}$ within the confined space $(-\infty, \rho_{\max}]^N$ while $\bm{V}$ and $\bm{\Theta}$ are fixed.
\end{itemize}

\begin{Proposition}\label{MP-PA-VAT-convergence}
    The MP-PA-VAT algorithm is an iterative improvement algorithm and converges.
\end{Proposition}

\noindent The proof is similar to that of Proposition \ref{BS-PA-VAT-convergence} and is omitted.

\subsection{Generalization to Soft Max-Min SINR}\label{SM-Subsection}

\subsubsection{Performance Function}\label{SM-Objective}
The soft max-min performance function is formulated as:
\begin{align}\label{soft-max-min-objective}
    \Phi_{{\sf SM}}(\bm{V}, \mathbf{\Theta}, \bm{\rho}) &=  \sum_{n=1}^{N} \int_{V_n} \gamma^{(n)}_{{\sf SM}} (\bm{q}; \mathbf{\Theta}, \bm{\rho}) \lambda(\bm{q}) d\bm{q}, \\
    \textrm{s.t. } \rho_n &\leq \rho_{\max} \qquad \forall n \in \{1,\cdots, N\}, \label{soft-max-min-objective-constraint}
\end{align}
where 
\begin{align}\label{Soft-max-min}
\gamma^{(n)}_{\sf SM}(\bm{q}; \mathbf{\Theta}, \bm{\rho})
= - \exp  \Bigg[ \frac{\alpha}{\big(\mathtt{SINR^{(n)}_{lin}}(\bm{q}; \mathbf{\Theta}, \bm{\rho}) + \nu \big)^{\xi}}\Bigg].
\end{align}

The hyperparameter $\alpha$ controls the softness of the max-min policy, with larger values resulting in the domination of the smallest SINR in $\gamma_{{\sf SM}}$. Thus, for large $\alpha$ values, optimizing $\Phi_{{\sf SM}}$ reduces to maximizing the minimum SINR. 
Conversely, smaller $\alpha$ values involve more SINR values in the performance function. To prevent users with very low SINR from dominating the performance, a small offset parameter $\nu$ is introduced. 
The exponent $\xi \leq 1$ compresses the dynamic range and enhances performance in the high-SINR regime.

\subsubsection{Optimal Configuration}\label{SM-Optimal-Configuration}

As before, we derive the necessary optimality conditions for each variable while holding the other two variables fixed to optimize the performance function $\Phi_{{\sf SM}}(\bm{V}, \mathbf{\Theta}, \bm{\rho})$ over cell partitioning, BS antenna tilts, and transmission powers.
\begin{Proposition}\label{optimal-cell-partitioning-SM-performance}
    The optimal cell partitioning $\bm{V}^*(\bm{\Theta}, \bm{\rho}) = \left(V_1^*(\bm{\Theta}, \bm{\rho}), \cdots, V_N^*(\bm{\Theta}, \bm{\rho}) \right)$ that maximizes the performance function $\Phi_{{\sf SM}}$ for a given $\bm{\Theta}$ and $\bm{\rho}$ is given by:
\begin{multline}\label{optimal-cell-partitioning-SM}
   \!\!\!\!\! V_n^*(\bm{\Theta}, \bm{\rho}) = \big\{\bm{q} \in Q \mid \mathtt{RSS_{dBm}^{(n)}}(\bm{q}; \theta_n, \rho_n) \geq \mathtt{RSS_{dBm}^{(k)}}(\bm{q}; \theta_k, \rho_k), \\ \textrm{ for all } 1 \leq k \leq N \big\},
\end{multline}  
for each $n \in \{1, \cdots, N\}$. The ties can be broken arbitrarily.
\end{Proposition}
\textit{Proof. }
Since $\alpha$, $\nu$, and $\xi$ are constants, $\gamma_{\sf SM}^{(n)}(\bm{q}; \mathbf{\Theta}, \bm{\rho}) \geq \gamma_{\sf SM}^{(k)}(\bm{q}; \mathbf{\Theta}, \bm{\rho})$ for all $k \neq n$ is the same as Eq. \eqref{optimal-cell-partitioning-SINR-proof} in Appendix \ref{Appendix_A}; therefore, the rest of the proof follows from that of Proposition \ref{optimal-V-SINR}. $\hfill\blacksquare$

\begin{figure*}[t!]
    \begin{multline}\label{gradient-vector-theta-SM-equation}
    \frac{\partial \Phi_{{\sf SM}}}{\partial \theta_n} =   \int_{V_n }  \frac{2.4 \log 10 \cdot \alpha \, \xi \,\left(\theta_{n,\bm{q}} - \theta_n \right) \SINR^{(n)}_{\mathtt{lin}} }{\left(\SINR^{(n)}_{\mathtt{lin}} + \nu\right)^{\xi+1} \cdot \theta^2_{\text{3dB}}} 
    \cdot \exp \! \left[ \frac{\alpha}{(\SINR^{(n)}_{\mathtt{lin}} + \nu )^{\xi}}\right]   \lambda(\bm{q})d\bm{q}   \\
    - \sum_{i\neq n}   \int_{V_i }  \frac{\alpha \, \xi \, \SINR^{(i)}_{\mathtt{lin}}}{\left(\SINR^{(i)}_{\mathtt{lin}} + \nu\right)^{\xi+1}} 
    \cdot \exp \! \left[ \frac{\alpha}{(\SINR^{(i)}_{\mathtt{lin}} + \nu )^{\xi}}\right] \times  
    \left[ \frac{2.4 \log 10}{\theta^2_{\text{3dB}}} \cdot \left(\theta_{n,\bm{q}} - \theta_n \right) \cdot {\frac{\RSS^{(n)}_{\mathtt{lin}}}{\sum_{j\neq i}^{} \RSS^{(j)}_{\mathtt{lin}} + \sigma^2_{\mathtt{lin}}}} \right]   \lambda(\bm{q})d\bm{q}.    
\end{multline}
\end{figure*}

\begin{figure*}[t!]
    \begin{multline}\label{gradient-SM-wrt-power-equation}
    \frac{\partial \Phi_{{\sf SM}}}{\partial \rho_n} =   \int_{V_n } \frac{\alpha\xi \cdot \mathtt{SINR_{\mathtt{lin}}^{(n)}} \cdot \frac{\ln(10)}{10} }{\big(\mathtt{SINR_{\mathtt{lin}}^{(n)}} + \nu\big)^{\xi + 1}} \cdot e^{\frac{\alpha}{\big(\mathtt{SINR^{(n)}_{\mathtt{lin}}} + \nu \big)^\xi}}   \lambda(\bm{q})d\bm{q}   
    - \sum_{i\neq n}   \int_{V_i } \frac{\alpha\xi\cdot \big[\mathtt{SINR_{\mathtt{lin}}^{(i)}} \big]^2 \cdot \frac{\ln(10)}{10}\cdot \mathtt{RSS_{\mathtt{lin}}^{(n)}}    }{\big(\mathtt{SINR_{\mathtt{lin}}^{(i)}} + \nu\big)^{\xi + 1}\cdot \mathtt{RSS_{\mathtt{lin}}^{(i)}}} \cdot e^{\frac{\alpha}{\big(\mathtt{SINR_{\mathtt{lin}}^{(i)}} + \nu \big)^\xi}}   \lambda(\bm{q})d\bm{q}.
\end{multline}
\end{figure*}

The following proposition provides the gradient of the performance function w.r.t. the vertical antenna tilts.
\begin{Proposition}\label{gradient-vector-theta-SM}
    The partial derivative of $\Phi_{{\sf SM}}$ w.r.t. the BS vertical antenna tilt $\theta_n$ is given by Eq. (\ref{gradient-vector-theta-SM-equation}), on top of the next page,
where $\SINR^{(n)}_{\mathtt{lin}}(\bm{q}; \bm{\Theta}, \bm{\rho})$, $\RSS^{(n)}_{\mathtt{lin}}(\bm{q}; \theta_n, \rho_n)$, and $V_n(\bm{\Theta}, \bm{\rho})$ are written as $\SINR^{(n)}_{\mathtt{lin}} $, $\RSS^{(n)}_{\mathtt{lin}}$, and $V_n$, respectively, for the brevity of notation.
\end{Proposition}
\noindent The proof is similar to that of Proposition \ref{gradient-equation-SINR} and is omitted.

Next, we provide the expression for the gradient of $\Phi_{{\sf SM}}$ w.r.t. the BS transmission powers $\bm{\rho}$.

\begin{Proposition}\label{gradient-SM-wrt-power}
    The partial derivative of $\Phi_{{\sf SM}}$ w.r.t. the BS transmission power $\rho_n$ is given by Eq. (\ref{gradient-SM-wrt-power-equation}), on top of the next page,
where $\SINR^{(n)}_{\mathtt{lin}} $, $\RSS^{(n)}_{\mathtt{lin}}$, and $V_n$ are shorts for $\SINR^{(n)}_{\mathtt{lin}}(\bm{q}; \bm{\Theta}, \bm{\rho})$, $\RSS^{(n)}_{\mathtt{lin}}(\bm{q}; \theta_n, \rho_n)$, and $V_n(\bm{\Theta}, \bm{\rho})$, respectively, for the brevity of notation.
\end{Proposition}
\noindent The proof is similar to that of Proposition \ref{sinr-power-allocation-gradient} and is omitted.

\subsubsection{Proposed Algorithm}\label{SM-Algorithm}
Using Propositions \ref{optimal-cell-partitioning-SM-performance}, \ref{gradient-vector-theta-SM}, and \ref{gradient-SM-wrt-power}, we design an alternating optimization algorithm, called the soft max-min power allocation and vertical antenna tilt (SMM-PA-VAT) optimization algorithm, similar to Algorithm \ref{BS_PA_VAT_Algorithm}.

\begin{Proposition}\label{SMM-PA-VAT-convergence}
    The SMM-PA-VAT algorithm is an iterative improvement algorithm and converges.
\end{Proposition}
\noindent The proof resembles the one of Proposition \ref{BS-PA-VAT-convergence} and  is omitted.

\section{Case Study}\label{case-study}

To evaluate the effectiveness and performance of the theoretical frameworks proposed, simulations were conducted using a practical case study. The subsequent sections begin by introducing the network configuration. Then, the numerical optimization results are presented, followed by a generalization to the case of probabilistic line-of-sight (LoS) condition for ground user links.

\subsection{Network Setup}\label{experimental-setup}

\subsubsection{Deployment Setup}\label{Deployment-Setup}
\begin{figure*}[!t]
\centering
\subfloat[GUEs, setup as per Sec.~\ref{experimental-setup}.]{\includegraphics[width=56mm]{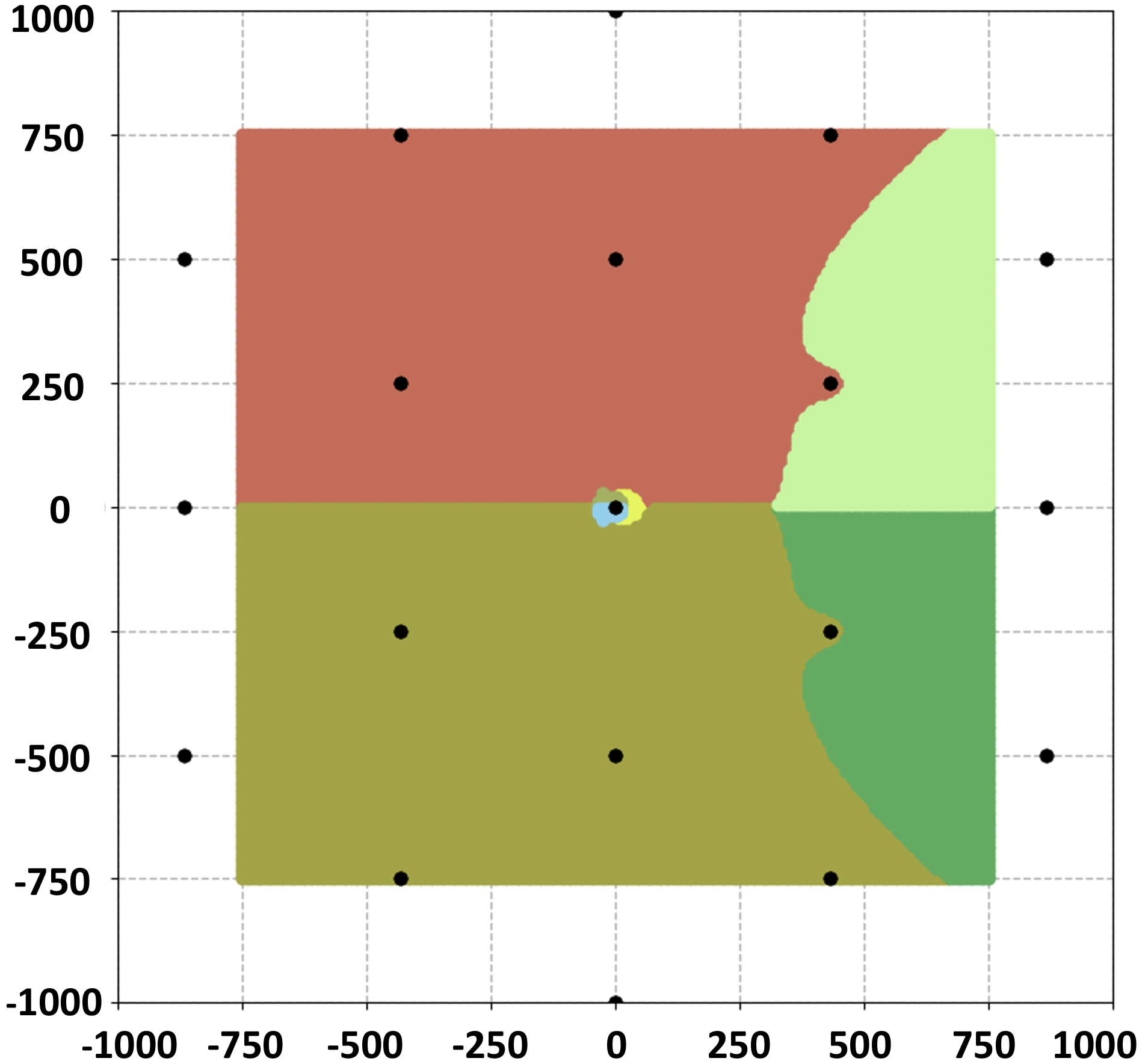}
\label{TWC-Normal-Size-GUE}}
\hspace*{3mm}
\subfloat[GUEs, setup as per Section~\ref{experimental-setup} $\times$2.]{\includegraphics[width=56mm]{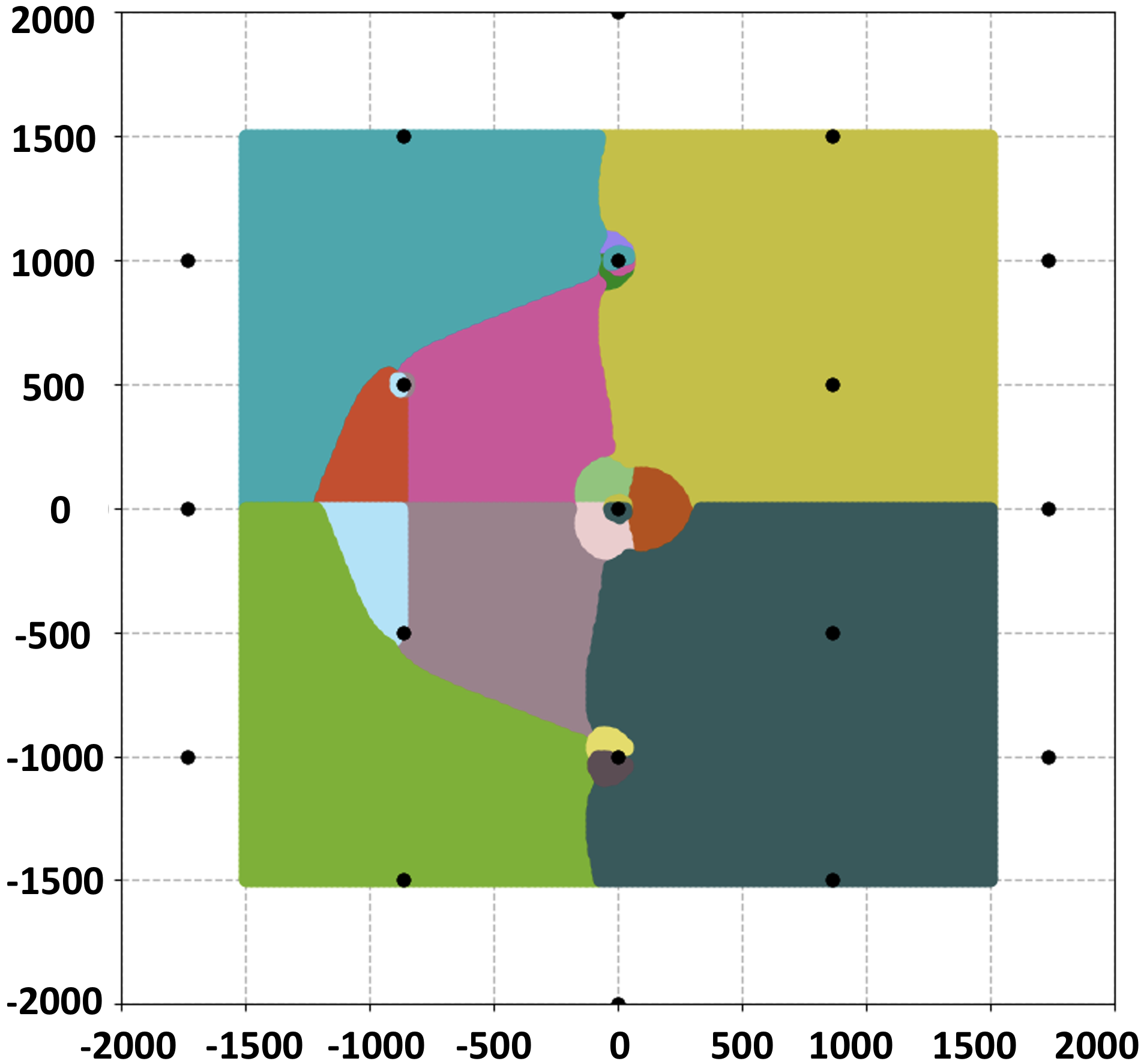}
\label{TWC-Large-Size-GUE}}
\hspace*{3mm}
\subfloat[GUEs, setup as per Section~\ref{experimental-setup} $\times$4.]{\includegraphics[width=56mm]{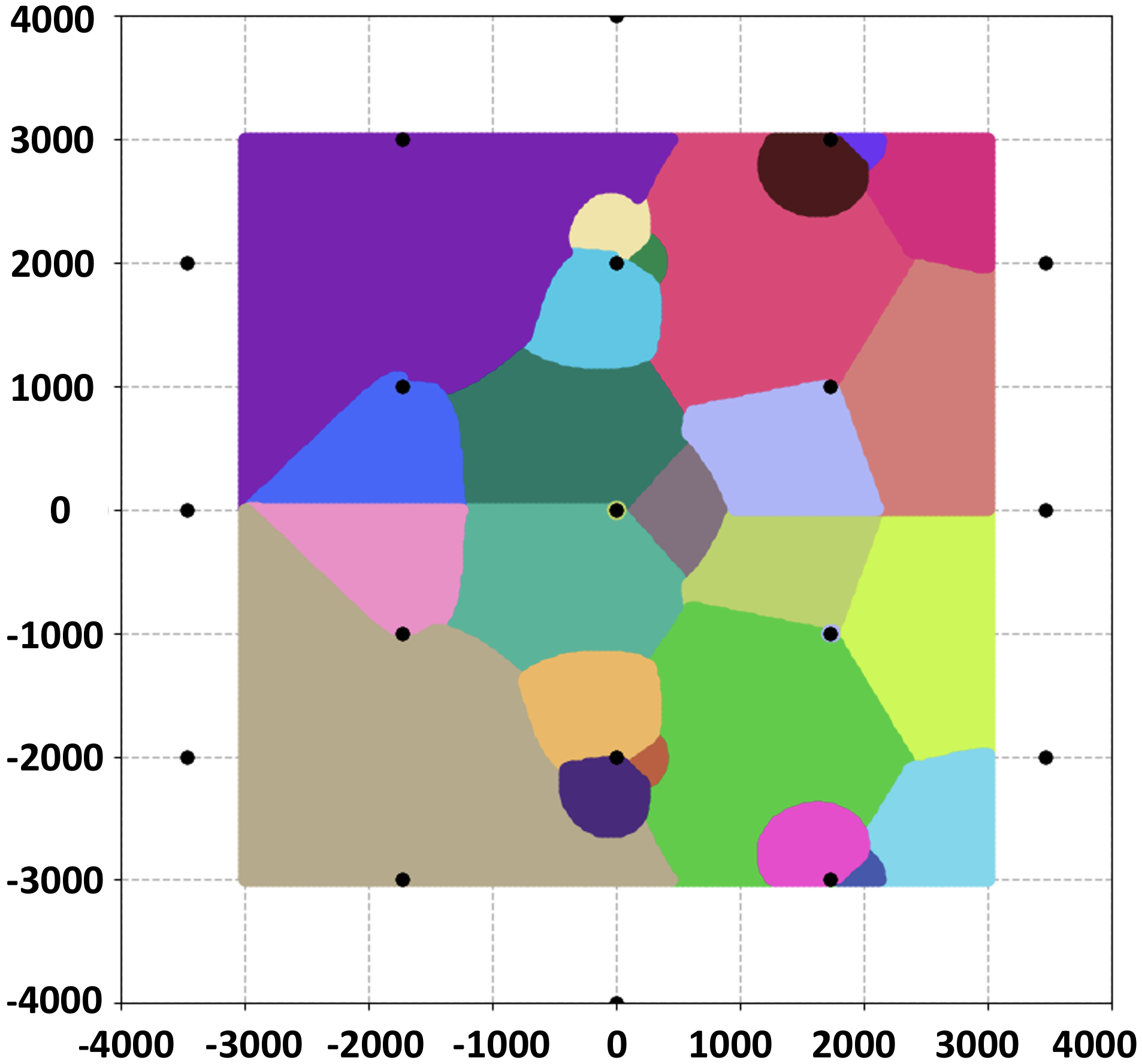}
\label{TWC-Very-Large-Size-GUE}}
\\\vspace*{3mm}
\subfloat[UAVs, setup as per Section~\ref{experimental-setup}.]{\includegraphics[width=56mm]{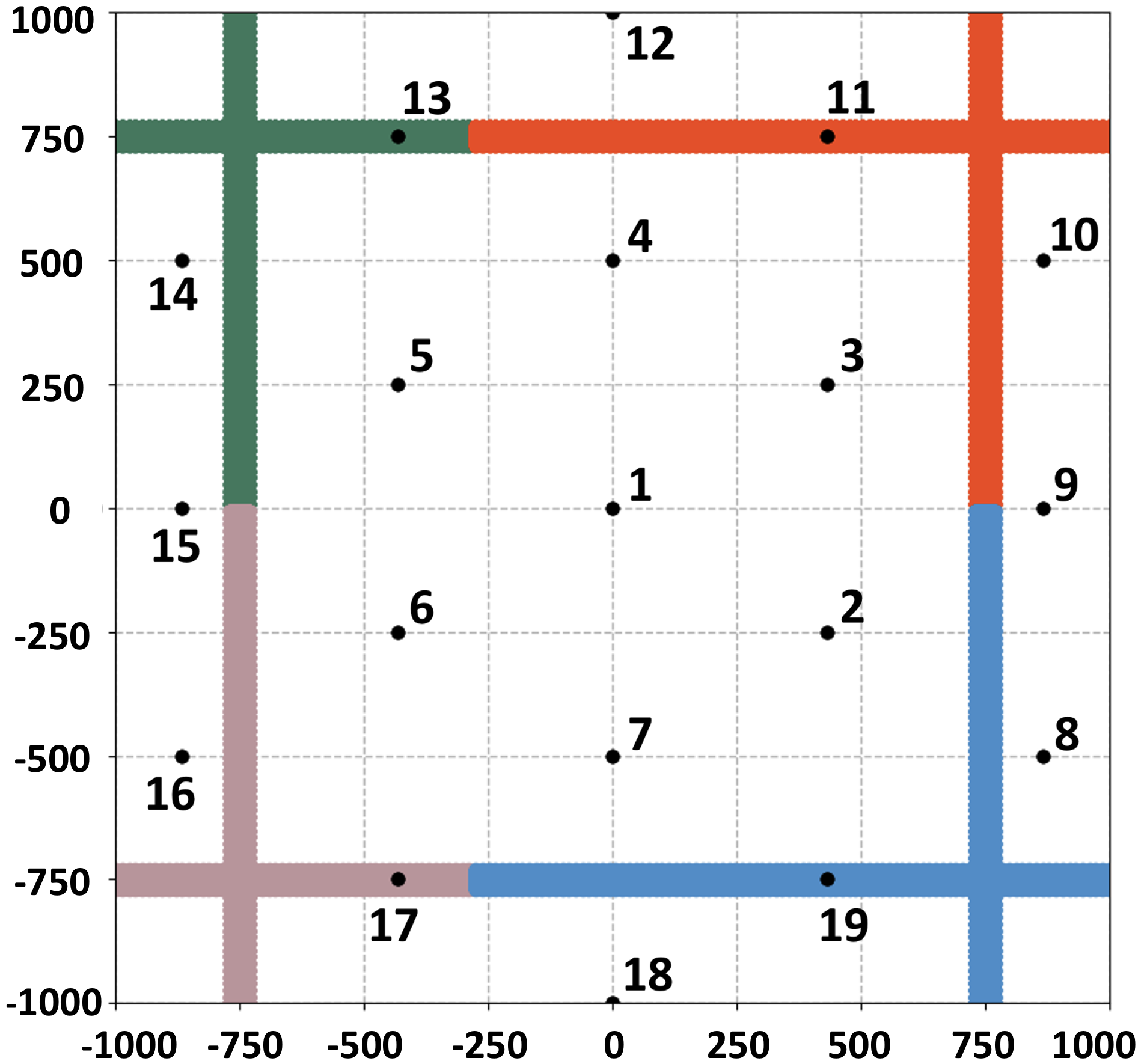}
\label{TWC-Normal-Size-UAV}}
\hspace*{3mm}
\subfloat[UAVs, setup as per Section~\ref{experimental-setup} $\times$2.]{\includegraphics[width=56mm]{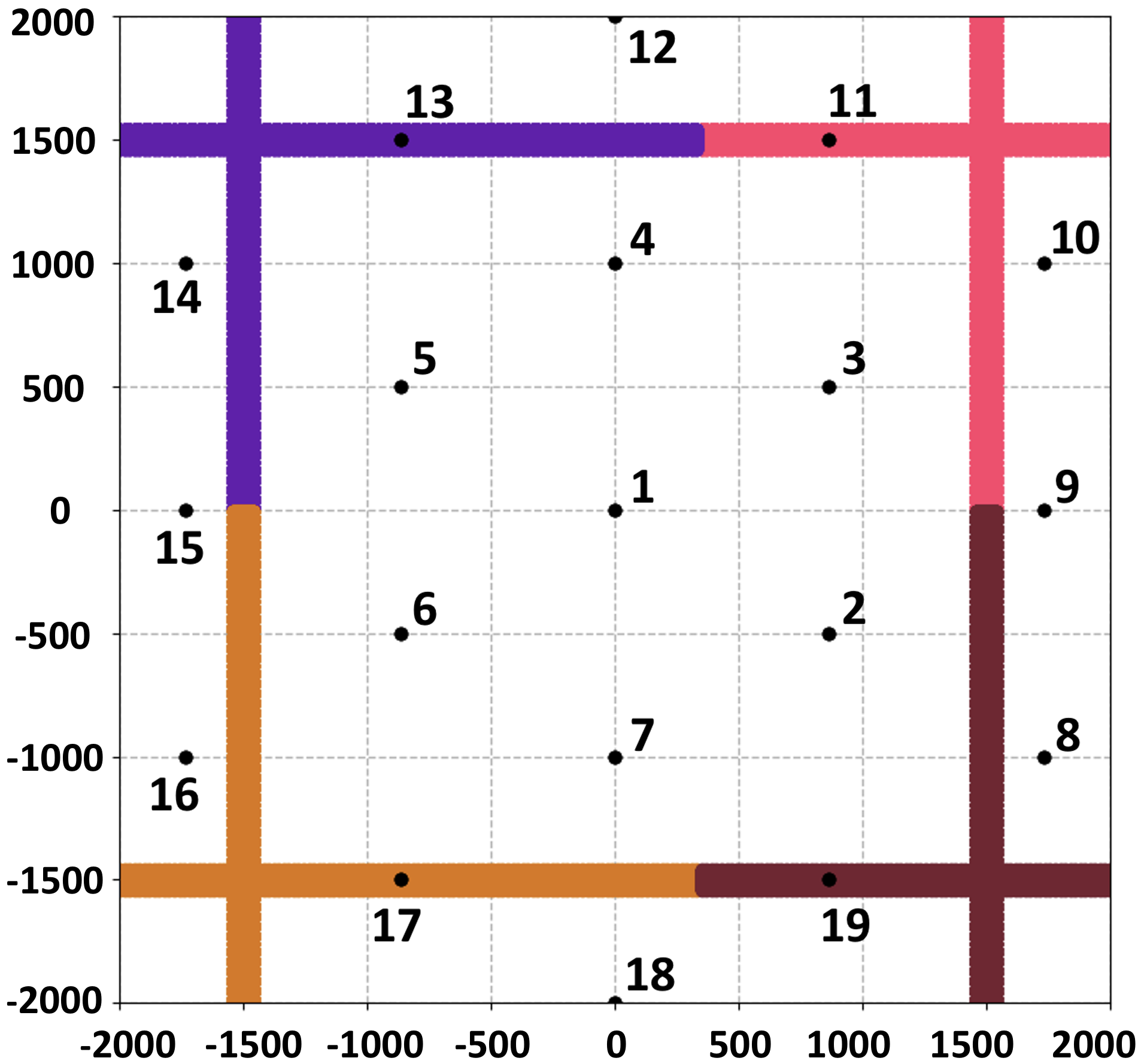}
\label{TWC-Large-Size-UAV}}
\hspace*{3mm}
\subfloat[UAVs, setup as per Section~\ref{experimental-setup} $\times$4.]{\includegraphics[width=56mm]{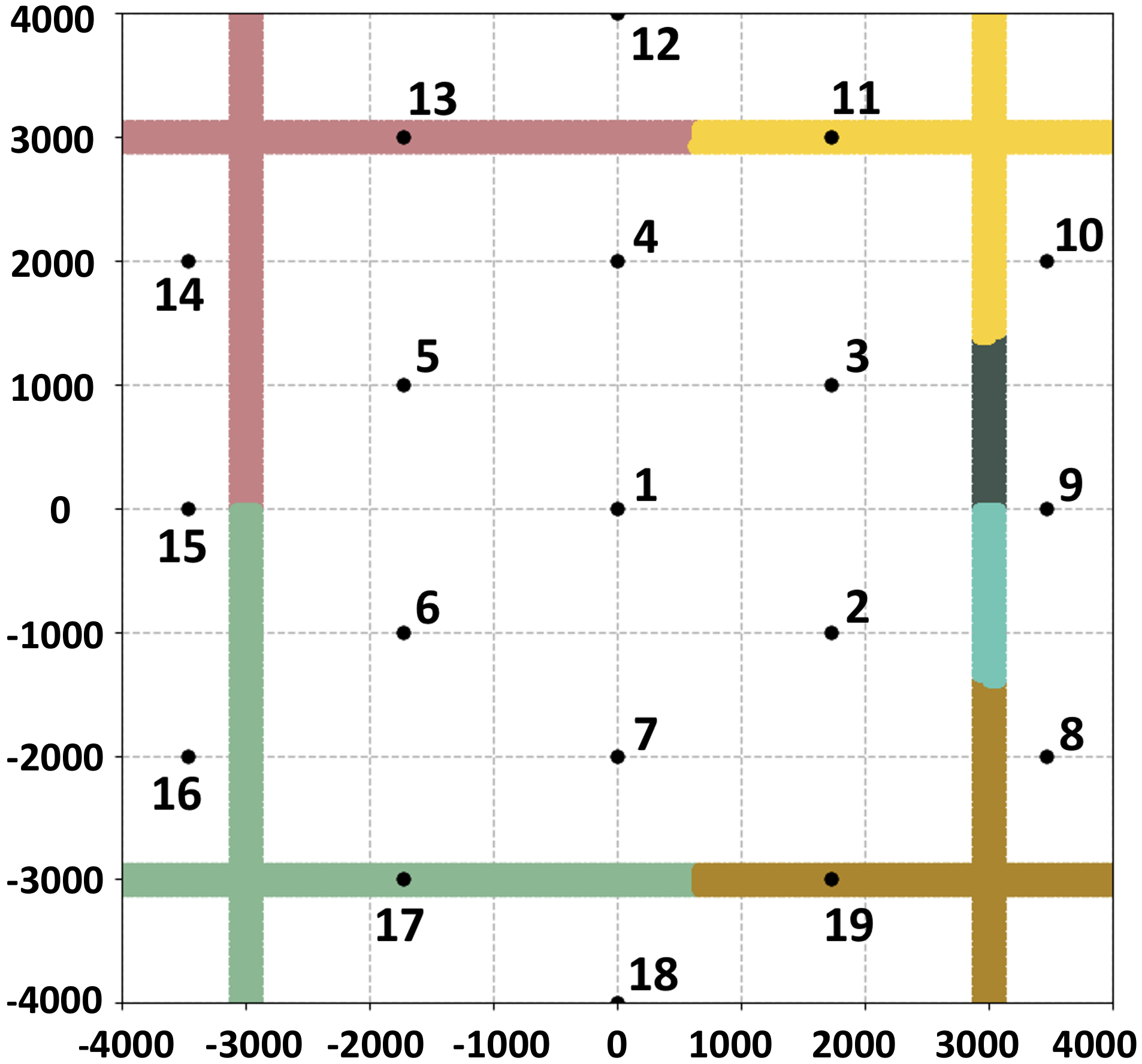}
\label{TWC-Very-Large-Size-UAV}}
\captionsetup{justification=justified}
\caption{Optimized GUEs and UAVs cell partitioning for the Max-SINR-PA-VAT algorithm with $r = 0.5$. Simulations are carried out for three different target region sizes and BS intersite distances.}
\label{TWC-Cell-Partitionings}
\end{figure*}

We examine a practical cellular network that consists of $19$ sites arranged in a hexagonal layout, where the inter-site distance (ISD) is $500$ meters. The configuration of this network and the BS deployment site indices are depicted in Fig.~\ref{TWC-Normal-Size-UAV}. Each site, here denoted by $k$, is associated with three sectors or cells. These cells have BSs located at the same positions (denoted by vector $\bm{p}_{3\times k-2} = \bm{p}_{3\times k-1} = \bm{p}_{3\times k}$), but they have different azimuth orientations. Specifically, the azimuth orientations are $\phi_{3\times k-2} = 0^\circ, \phi_{3\times k-1} = 120^\circ\textrm{, and } \phi_{3\times k} = 240^\circ$. Hence, a total of $N=57$ BSs are present, each requiring optimization of its vertical antenna tilt and transmission power values. All BSs share a common height of $h_{n,\mathrm{B}} = 25$m for $n$ ranging from $1$ to $N=57$. The maximum transmission power allowed for all BSs is $43$\,dBm.

The ground users are spatially distributed across a square area  $Q_G = [-750, 750] \times [-750, 750]$ as shown in Fig. \ref{TWC-Normal-Size-GUE}. Their distribution follows a uniform density function $\lambda_\textrm{G}(\bm{q})$ and they are assumed to have a fixed height of $h_\textrm{G} = 1.5$\,m. The UAVs are distributed over four vertical aerial corridors, represented by $Q_U = Q_1\cup Q_2\cup Q_3 \cup Q_4$, following a uniform density function $\lambda_\textrm{U}(\bm{q})$. These corridors, illustrated in Fig. \ref{TWC-Normal-Size-UAV}, are defined as $Q_1 = [-770, -730]\times [-1000, 1000]$, $Q_2 = [-1000, 1000]\times [-770, -730]$, $Q_3 = [-1000, 1000]\times [730, 770]$, and $Q_4 = [730, 770]\times [-1000, 1000]$. The heights of the corridors are set to $h_1 = h_4 = 150$\,m and $h_2 = h_3 = 120$\,m. The overall density function $\lambda(\bm{q})$, which represents the user distribution in $Q = Q_G \cup Q_U$, is a mixture of $\lambda_\textrm{G}(\bm{q})$ and $\lambda_\textrm{U}(\bm{q})$. Specifically, $\lambda(\bm{q}) = r\lambda_\textrm{G}(\bm{q}) + (1-r) \lambda_\textrm{U}(\bm{q})$, where $r$ is the mixing ratio. Throughout the study, we consider three different values for the parameter $r$, namely $1$, $0$, and $0.5$. These values correspond to optimizing the cellular network exclusively for ground users, exclusively for UAVs, and for both ground users and UAVs with equal priority, respectively.

\subsubsection{Channel Setup}\label{Channel-Setup}

According to the specifications provided by 3GPP \cite{3GPP36777,3GPP38901}, for a carrier frequency of 2GHz and under line-of-sight conditions, the values of $a_{\bm{q}}$ and $b_{\bm{q}}$ are set as follows:
\begin{equation}\label{a_q_values}
a_{\bm{q}} =
\begin{cases}
34.02\,\textrm{dB}, & \text{if}\ \bm{q}\in Q_U, \\
38.42\,\textrm{dB}, & \text{if}\ \bm{q}\in Q_G,
\end{cases}
\end{equation}
\begin{equation}\label{b_q_values}
b_{\bm{q}} =
\begin{cases}
22 \textrm{ (for a pathloss exponent of 2.2)}, & \text{if}\ \bm{q}\in Q_U, \\
30 \textrm{ (for a pathloss exponent of 3.0)}, & \text{if}\ \bm{q}\in Q_G.
\end{cases}
\end{equation}

Furthermore, the directional antennas have the vertical half-power beamwidth of $\theta_\textrm{3dB} = 10^{\circ}$, the horizontal half-power beamwidth of $\phi_\textrm{3dB} = 65^{\circ}$, and the maximum antenna gain of $A_\textrm{max} = 14$\,dBi at the boresight.

\subsection{Experimental Results}\label{Experimental-Results}

Each of the proposed algorithms is initialized with a random cell partitioning where each user at location $\bm{q}\in Q$ is assigned to a BS in a random manner. Additionally, the initial values of $\theta_n$ for all $n \in {1, \cdots, N}$ are set to $0^\circ$. In the case of the BS-VAT algorithm, which optimizes RSS  across network users, all $\rho_n$ values are set at a fixed level of $43$\,dBm. This power value is chosen because it is the straightforward optimal transmission power in the absence of interference. Conversely, for all other algorithms, the initial values of all $\rho_n$ are initialized to $0$\,dBm. The learning rate $\eta_0$ and the constant $\kappa$ are set as 0.01 and 0.999, respectively. Finally, the convergence error thresholds, $\epsilon_1$, $\epsilon_2$, and $\epsilon_3$ are chosen to be $10^{-8}$.

\begin{figure}[!t]
\centering
\subfloat[Max-RSS-VAT algorithm.]{\includegraphics[width=\figwidth]{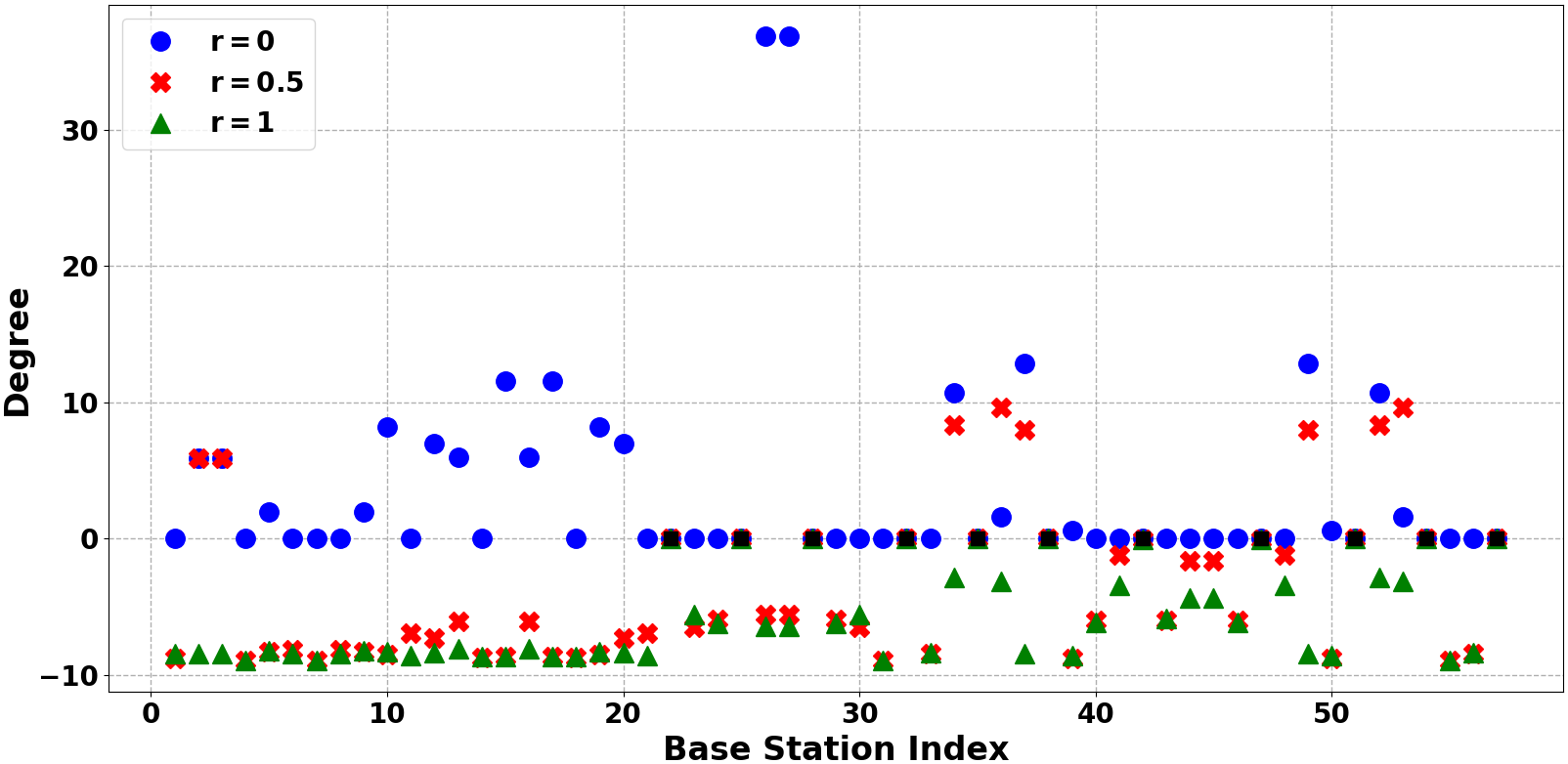}
\label{TWC-RSS-Optimal-Theta}}
\hspace{0mm}\\
\vspace*{3mm}
\subfloat[MP-PA-VAT algorithm with $\mu =\nu = 0.1$.]{\includegraphics[width=\figwidth]{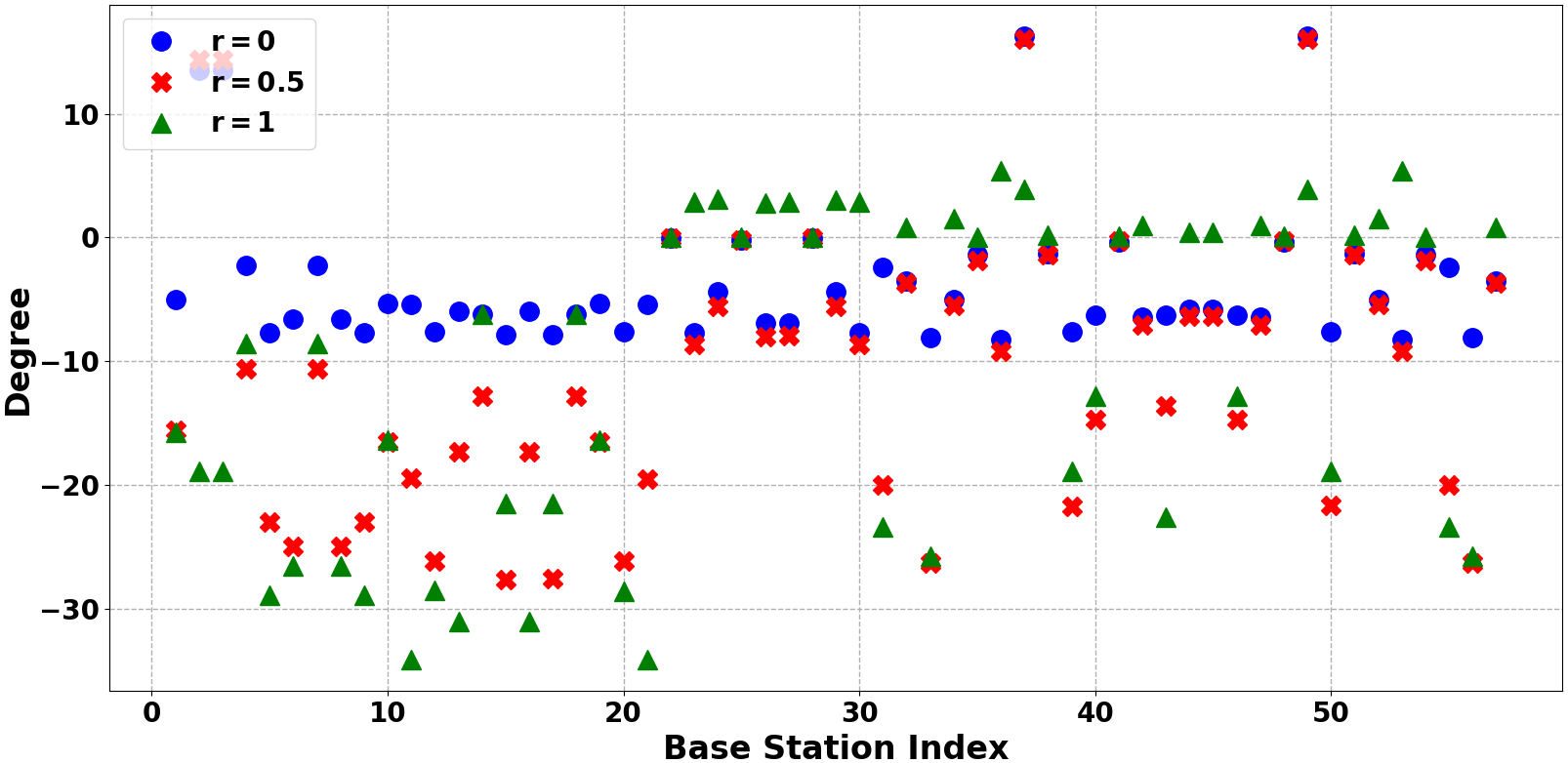}
\label{TWC-MP-010-Optimal-Theta}}
\captionsetup{justification=justified}
\caption{Optimized vertical tilts,  $\theta_i^*$: (a) Max-RSS-VAT and (b) MP-PA-VAT Algorithm with $\mu =\nu = 0.1$. Optimized for: GUEs only (green triangles, $r=1$), UAVs only (blue circles, $r=0$), and both GUEs and UAVs (red crosses, $r=0.5$).}
\label{TWC-Optimal-Theta}
\end{figure}

\subsubsection{Optimal Vertical Antenna Tilts}
Figs. \ref{TWC-RSS-Optimal-Theta} and \ref{TWC-MP-010-Optimal-Theta} display the optimal vertical antenna tilts,  $\theta_n^*$, for the Max-RSS-VAT and the MP-PA-VAT algorithms. Each figure showcases the optimal tilts for three scenarios: $r = 0$ (represented by blue circles), $r = 0.5$ (depicted by red crosses), and $r = 1$ (illustrated by green triangles). 
As anticipated, in the Max-RSS-VAT algorithm, where the BS antenna tilts are configured to optimize the average RSS across network users, prioritizing the optimization process for either ground users ($r=1$) or UAVs ($r=0$) leads to all BSs to be either downtilted or uptilted, respectively. This outcome stems from the fact that interference effects are not taken into account when the objective is to maximize the average RSS. However, in the case of $r = 0.5$, a tradeoff is achieved, resulting in a combination of uptilted and downtilted antennas. 
When it comes to the MP-PA-VAT algorithm, adjusting the vertical antenna tilts to optimize the system for any of the three scenarios ($r = 0$, $r = 0.5$, and $r = 1$) leads to a combination of uptilted and downtilted base stations. Similar observations were made for the Max-SINR-PA-VAT and SMM-PA-VAT algorithms. This is due to the fact that interference plays a substantial role in influencing the performance functions of these algorithms. In addition, in certain network configurations, certain BSs may not have an impact on the performance functions. This situation is exemplified in Fig. \ref{TWC-RSS-Optimal-Theta}, where BSs that do not contribute to the performance function in any of the three simulated scenarios ($r = 0, 0.5,$ and $1$) are depicted as black squares.


\subsubsection{Optimal Transmission Power}
\begin{figure}[!t]
\centering
\subfloat[Max-SINR-PA-VAT algorithm.]{\includegraphics[width=\figwidth]{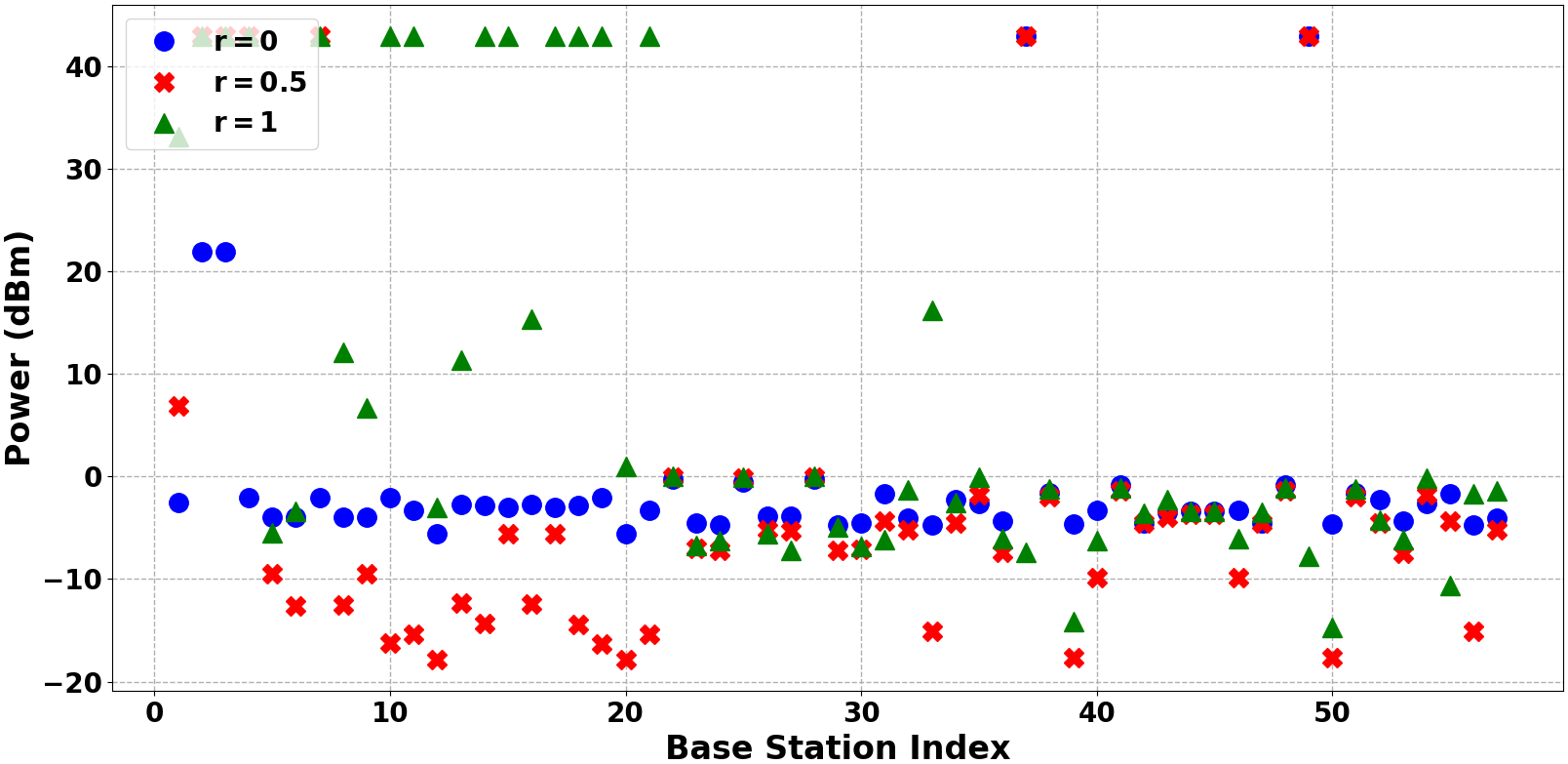}
\label{TWC-SINR-Optimal-Power}}
\hspace{0mm}\\
\vspace*{3mm}
\subfloat[MP-PA-VAT algorithm with $\mu =\nu = 0.1$.]{\includegraphics[width=\figwidth]{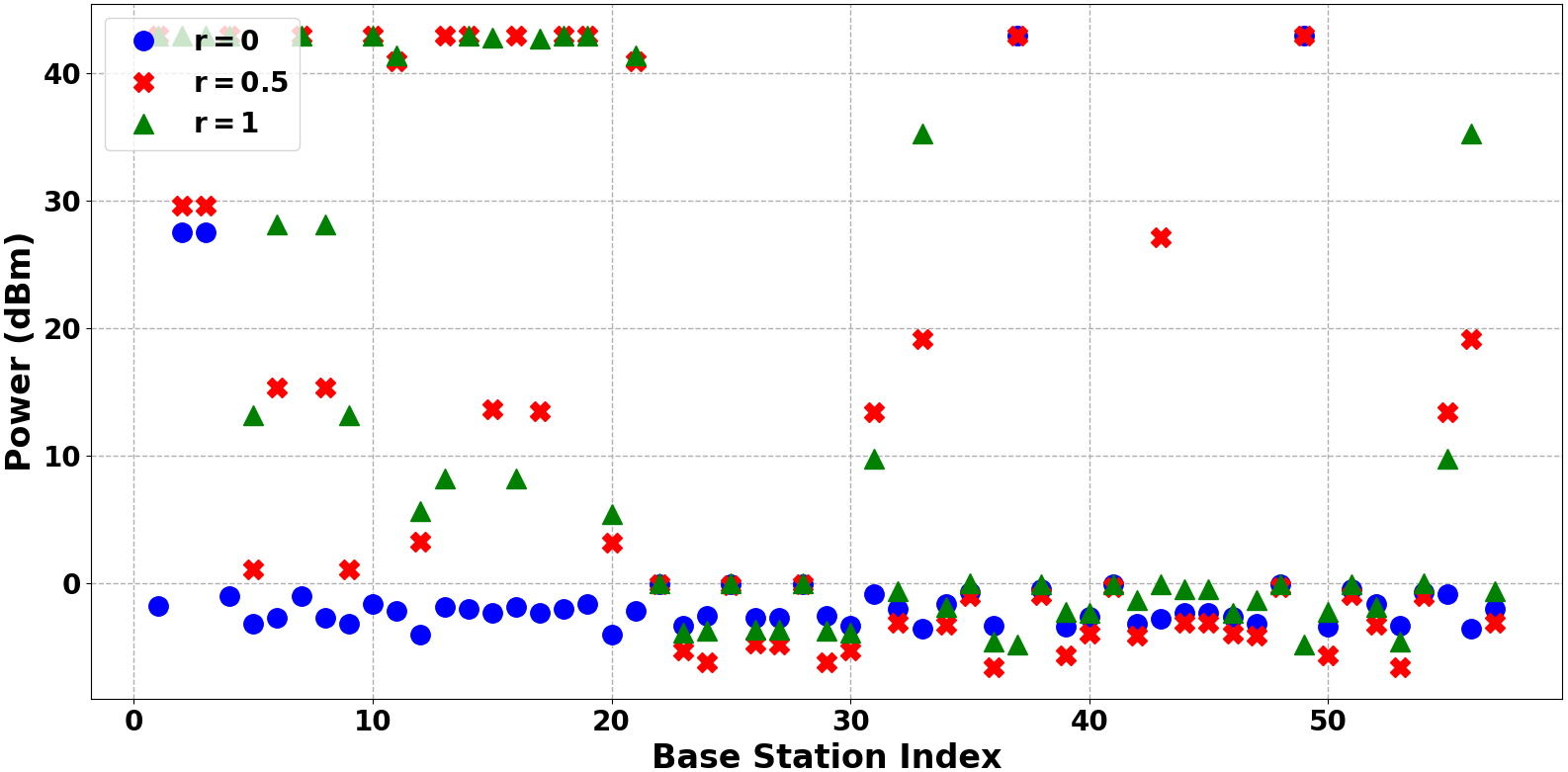}
\label{TWC-MP-010-Optimal-Power}}
\captionsetup{justification=justified}
\caption{Optimized transmission powers $\rho_i^*$ for: (a) Max-SINR-PA-VAT Algorithm and (b) MP-PA-VAT Algorithm with $\mu =\nu = 0.1$. Optimized for: GUEs only (green triangles, $r=1$), UAVs only (blue circles, $r=0$), and both GUEs and UAVs (red crosses, $r=0.5$).}
\label{TWC-Optimal-Power}
\end{figure}

Figs. \ref{TWC-SINR-Optimal-Power} and \ref{TWC-MP-010-Optimal-Power} present the optimal transmission power values,  $\rho_n^*$, for the Max-SINR-PA-VAT and MP-PA-VAT algorithms. In each of the three scenarios, namely $r=0$, $r=0.5$, and $r=1$, a subset of BSs operates at the maximum power level of $43$\,dBm, while another subset utilizes lower power levels, and the remaining BSs are deactivated. While not shown, similar observations are made for the SMM-PA-VAT algorithm. This is in contrast to the Max-RSS-VAT algorithm where all BSs are set to the optimal transmission power value of $43$\,dBm. 
However, as the target region and ISD grow larger, the impact of interference diminishes and more BSs become active. This is demonstrated in Fig.~\ref{TWC-Cell-Partitionings} where the Max-SINR-PA-VAT algorithm is utilized to determine the most favorable network configuration for three distinct combinations of GUE and UAV target region sizes and ISD values in the case of $r = 0.5$. 
The initial pair, depicted in Figs. \ref{TWC-Normal-Size-GUE} and \ref{TWC-Normal-Size-UAV}, corresponds to the setup described in Section \ref{experimental-setup}. For the second pair, showcased in Figs. \ref{TWC-Large-Size-GUE} and \ref{TWC-Large-Size-UAV}, the GUE target region, distance between UAV corridors, and their respective widths, along with the BS ISD, are all doubled. Consequently, the optimal partitioning of the GUE target region in Fig.~\ref{TWC-Large-Size-GUE} reveals an increased number of cells and more active BSs. Finally, expanding the setup in Figs. \ref{TWC-Large-Size-GUE} and \ref{TWC-Large-Size-UAV} by an additional factor of two yields the configuration depicted in Figs. \ref{TWC-Very-Large-Size-GUE} and \ref{TWC-Very-Large-Size-UAV}. The second expansion results in a further increase in the number of cells for both the optimal GUE and UAV target region partitioning. This is primarily due to the reduced impact of interference at larger distances, allowing more BSs to efficiently serve users in their vicinity.

\begin{figure}[!t]
\centering
\subfloat[Max-RSS-VAT algorithm.]{\includegraphics[width=\figwidth]{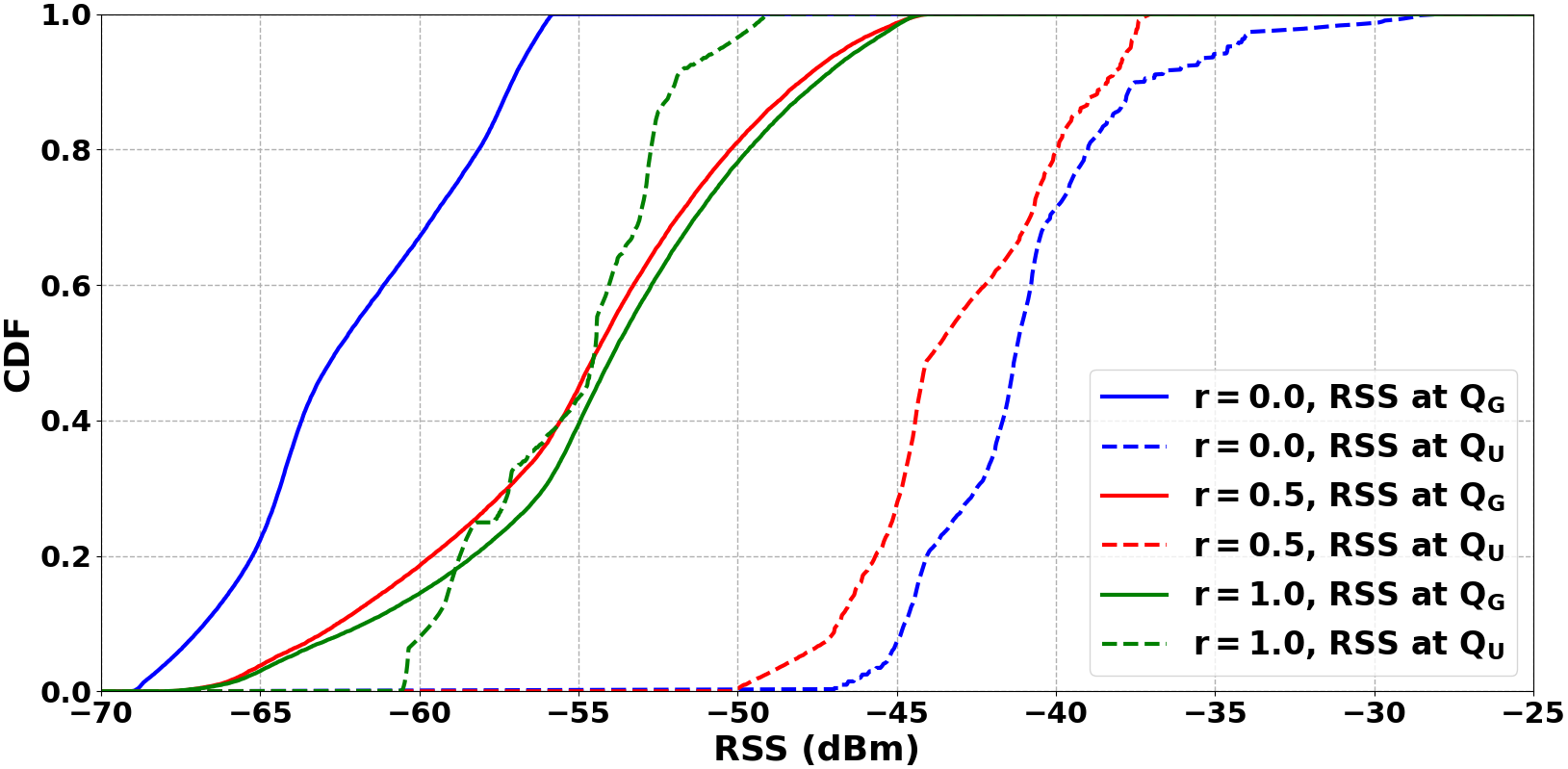}
\label{TWC-RSS-CDF-Plot}}
\hspace{0mm}\\
\vspace*{3mm}
\subfloat[MP-PA-VAT algorithm with $\mu =\nu = 0.1$.]{\includegraphics[width=\figwidth]{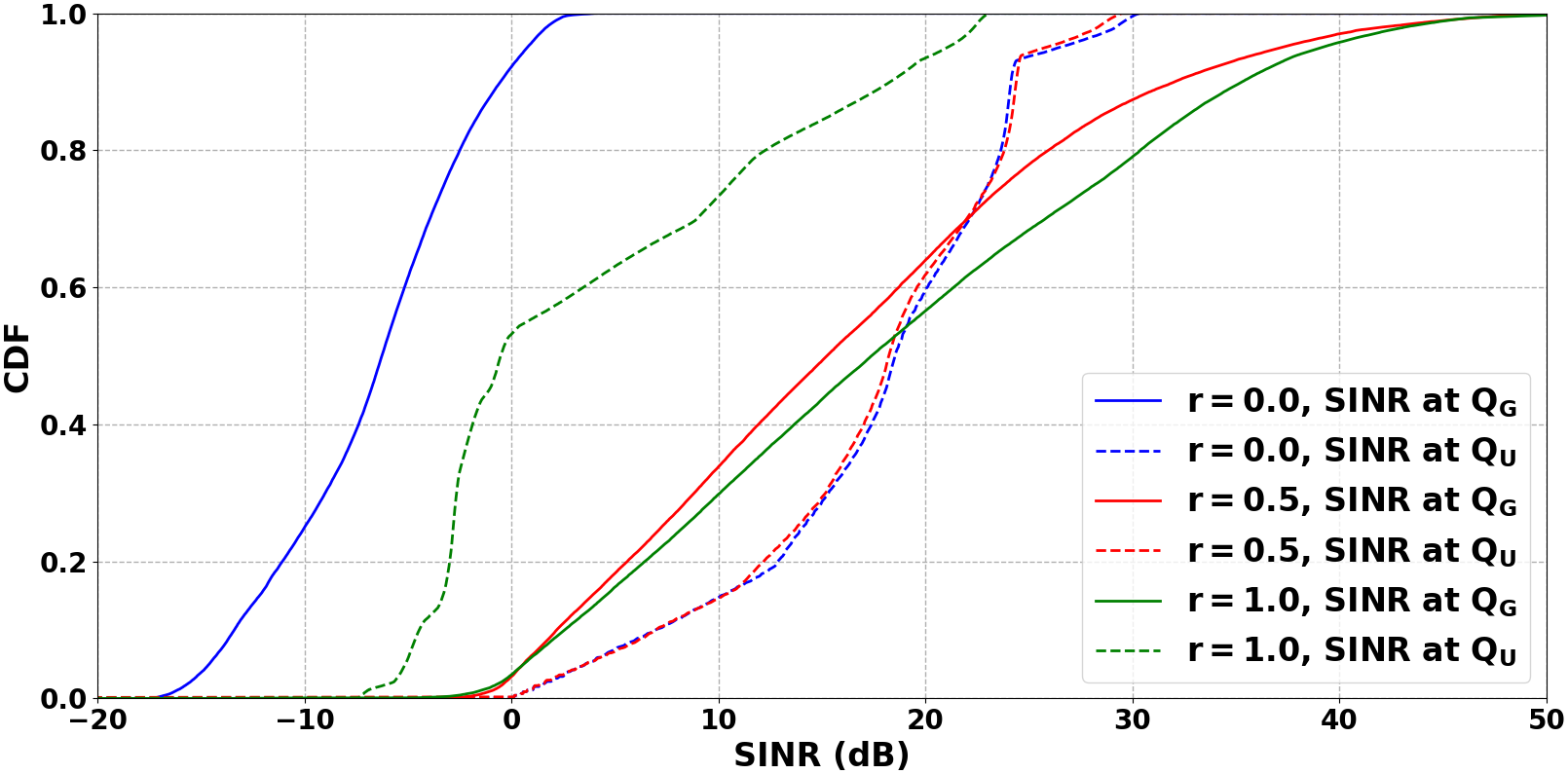}
\label{TWC-MP-010-CDF-Plot}}
\hspace{0mm}\\
\vspace*{3mm}
\subfloat[Max-SINR-PA-VAT algorithm under probabilistic LoS/NLoS.]{\includegraphics[width=\figwidth]{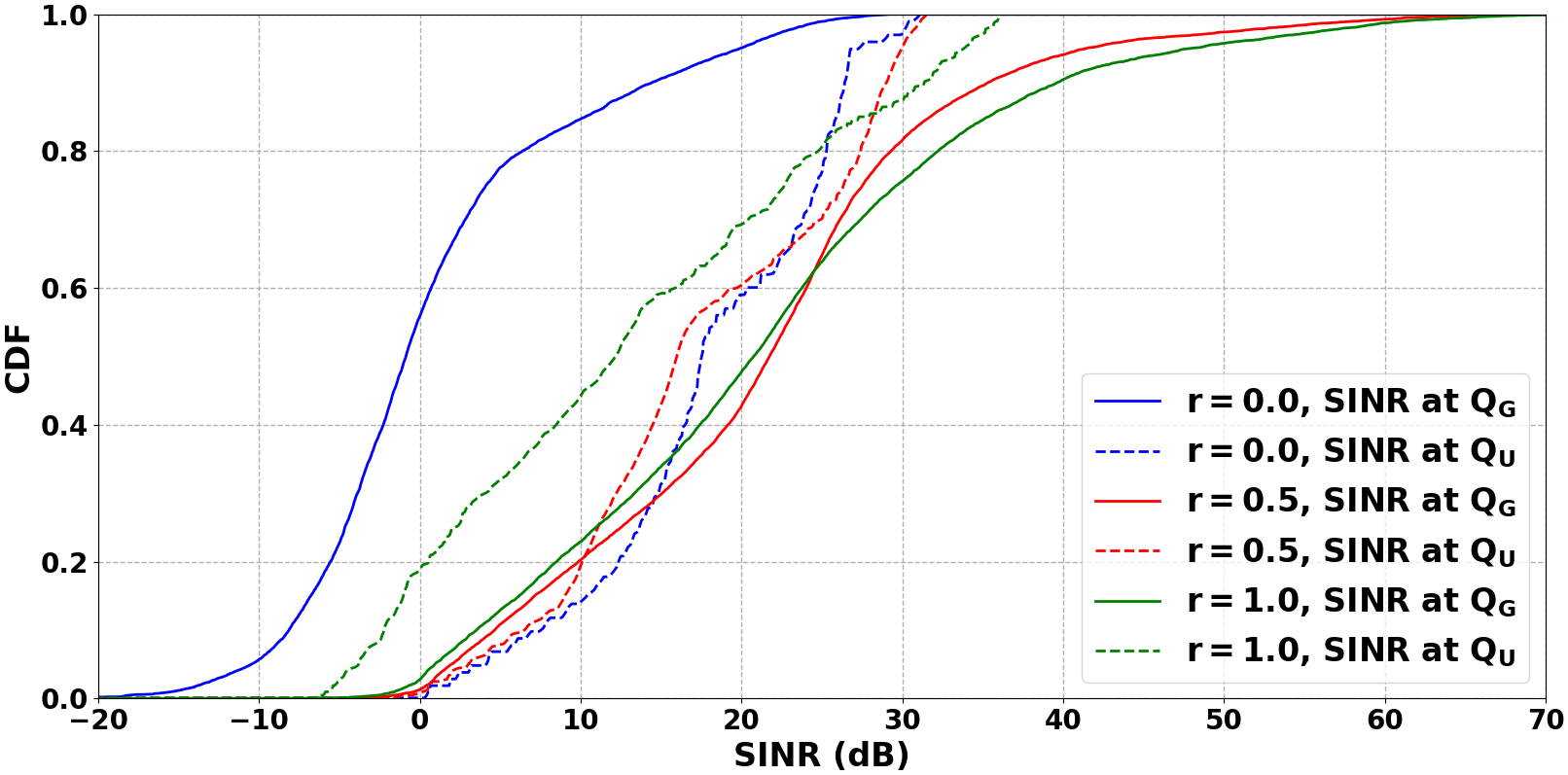}
\label{TWC-SINR-CDF-Plot-Prob-LoS}}
\captionsetup{justification=justified}
\caption{CDF of (a) the RSS for the Max-RSS-VAT algorithm, (b) the SINR for the MP-PA-VAT algorithm with $\mu =\nu = 0.1$, and (c) the SINR for the Max-SINR-PA-VAT algorithm under probabilistic LoS/NLoS condition. Dash-dash and solid curves represent UAVs and GUEs, respectively. Three optimization scenarios are shown: GUEs only ($r=1$), UAVs only ($r=0$), and both GUEs and UAVs ($r=0.5$).}
\label{TWC-CDF-Plots}
\end{figure}

\subsubsection{Performance Improvement}\label{Resulting-performance-improvement}


Fig. \ref{TWC-RSS-CDF-Plot} shows the cumulative distribution function (CDF) of the RSS perceived by ground users (solid line) and UAVs (dash-dash line) when the antenna tilts are optimized through the Max-RSS-VAT algorithm for ground users only ($r = 1$, green), UAVs only ($r = 0$, blue), and both ($r= 0.5$, red). Note that the ground user performance for $r = 1$ (green solid line) and the UAV performance for $r = 0$ (blue dash-dash line) can be regarded as respective upper bounds (in mean) since they entail optimizing all vertical tilts for ground users only and for UAVs only, respectively. Conversely, the ground user performance for $r = 0$ (blue solid line) and the UAV performance for $r = 1$ (green dash-dash line) can be regarded as respective baselines obtained when the vertical tilts are chosen ignoring ground users and UAVs, respectively. Fig. \ref{TWC-RSS-CDF-Plot} shows that for $r = 0.5$ the proposed Max-RSS-VAT algorithm reaches a satisfactory tradeoff by: (i) significantly boosting the RSS at UAVs (red dash-dash line) compared to the baseline (green dash-dash line) and approaching the upper bound (blue dash-dash line), and (ii) nearly preserving the RSS at ground users (red solid line) compared to the upper bound (green solid line). Specifically, the average RSS gain at UAVs amounts to $12$\,dB and comes at the expense of an average loss of only $0.7$\,dB at ground users.

Similarly, Fig. \ref{TWC-MP-010-CDF-Plot} shows the CDF of the SINR perceived by ground users and UAVs when antenna tilts and transmit power are optimized through the MP-PA-VAT algorithm with $\mu =\nu = 0.1$, for ground users only, UAVs only, and both. Fig. \ref{TWC-MP-010-CDF-Plot} shows that the proposed algorithm reaches an SINR tradeoff, boosting the average SINR at UAVs by $13$\,dB while only incurring an average loss of $2$\,dB at ground users. 

\subsection{Generalization to Probabilistic Line-of-Sight Conditions}\label{Probabilistic-LOS-NLOS}

While the channel setup used for simulations in Section \ref{Experimental-Results} assumed all GUEs to experience a non-line-of-sight (NLoS) condition, our framework is applicable to any given LoS and NLoS set up. Indeed, as per 3GPP channel modeling, the presence or absence of LoS conditions between a user at $\bm{q}$ and its corresponding BS only impacts the specific values of $a_{\bm{q}}$ and $b_{\bm{q}}$ for that particular user. Our framework is designed to accommodate generic values for these parameters. 
For instance, following the 3GPP model \cite{3GPP38901}, if the user location $\bm{q}$ is in the LoS of its corresponding BS, we can take that into account by changing the values of $38.42$\,dB and $30$ in Eqs. (\ref{a_q_values}) and (\ref{b_q_values}) to $34.02$\,dB and $22$, respectively.

Throughout this section, we update the notation from $a_{\bm{q}}$ and $b_{\bm{q}}$ to $a_{\bm{q}, n}$ and $b_{\bm{q}, n}$, respectively, to accommodate the presence or absence of LoS conditions between the user at $\bm{q}$ and BS $n$. In the remainder of this section, we assume that UAVs are consistently in a LoS condition because of their elevated altitude \cite{3GPP36777}; however, the same reasoning can also be applied to user locations $\bm{q} \in Q_U$. Let $\tau_{\bm{q}, n}$ be a binary label taking the value of $1$ if the user location $\bm{q}$ is in LoS with BS $n$ and $0$ otherwise. Then, we have: 
\begin{align}\label{a_qn_values}
a_{\bm{q},n} &=
\begin{cases}
34.02\,\textrm{dB}, & \text{if}\ \bm{q}\in Q_U, \\
34.02\,\textrm{dB}, & \text{if}\ \bm{q}\in Q_G \text{ and } \tau_{\bm{q},n} = 1, \\
38.42\,\textrm{dB}, & \text{if}\ \bm{q}\in Q_G \text{ and } \tau_{\bm{q},n} = 0,
\end{cases}
\\
b_{\bm{q},n} &=
\begin{cases}
22, & \text{if}\ \bm{q}\in Q_U, \\
22, & \text{if}\ \bm{q}\in Q_G \text{ and } \tau_{\bm{q},n} = 1, \\
30, & \text{if}\ \bm{q}\in Q_G \text{ and } \tau_{\bm{q},n} = 0.
\end{cases}
\end{align}
The pathloss in Eq. (\ref{eqn:Pathloss}) is then given by:
\begin{equation} \label{eqn:NewPathloss}
L_{n,\bm{q}} = a_{\bm{q}, n} + b_{\bm{q}, n} \log_{10}\left[\| \bm{q} - \bm{p}_n \|^2 + (h_{\bm{q}} -h_{n,\mathrm{B}})^2 \right]^{\frac{1}{2}},
\end{equation}
while all other notations remain unaltered and all propositions still hold.

A practical case study for probabilistic LoS conditions follows from the 3GPP standard guideline in which the probability of LoS between the GUE at $\bm{q} \in Q_G$ and BS $n$ located at $\bm{p}_n$ is given by:
\begin{align}\label{probability-LoS}
    \mathrm{Pr}_{\textrm{LoS}} = 
    \begin{cases}
    &\!\!\!\!\! 1,   \qquad\qquad\qquad\qquad\qquad\quad    \text{if}\ \| \bm{q} - \bm{p}_n \| \leq 18\textrm{m}, \\
    &\!\!\!\!\! \frac{18}{\| \bm{q} - \bm{p}_n \|} + \Big(1 - \frac{18}{\| \bm{q} - \bm{p}_n \|}\Big)e^{-\frac{\| \bm{q} - \bm{p}_n \|}{63}},   \quad\textrm{ } \text{otherwise}.
    \end{cases}
\end{align}
For each $\bm{q} \in Q_G$ and $n \in \{1, \cdots, N\}$, the label $\tau_{\bm{q}, n}$ is then created as follows: a scalar $u$ is sampled at random from the uniform distribution $u \sim \mathcal{U}[0, 1]$. The label $\tau_{\bm{q}, n}$ is set to $1$ if $u\leq \mathrm{Pr}_{\textrm{LoS}}$, and $0$ otherwise. Once labels are created, the Max-SINR-PA-VAT algorithm is executed for three scenarios: $r = 0$, $r = 0.5$, and $r = 1$. Fig. \ref{TWC-SINR-CDF-Plot-Prob-LoS} illustrates the CDF of the SINR experienced by GUEs and UAVs in the three different scenarios. Similar observations as the ones in Section~\ref{Resulting-performance-improvement} can be made from this figure, i.e., for $r = 0.5$, there is a tradeoff between optimizing GUE and UAV performance. This tradeoff results in a substantial overall improvement in the SINR perceived by UAVs without a severe degradation in the GUE SINR. 
This finding showcases the broad versatility of our framework and its ability to deal with varying link conditions between users and BSs. Moreover, the algorithms could potentially extract link conditions from existing radio coverage datasets, making our algorithms well-suited for diverse real-world applications.

\section{Conclusion}\label{conclusion}

In this paper, we took the first step towards creating a mathematical framework for optimizing antenna tilts and transmit power in cellular networks, with the goal of providing the best quality of service to both legacy ground users and UAVs flying along corridors. By applying quantization theory and designing iterative algorithms, we modeled realistic features of network deployment, antenna radiation patterns, and propagation channel models. Our proposed algorithms offer the capability to optimize coverage and signal quality 
while allowing for trade-offs between performance on the ground and along UAV corridors through adjustable hyperparameters. The optimal combinations of antenna tilts and transmit power, which are non-obvious and challenging to design, were shown to significantly enhance performance along UAV corridors. Importantly, these improvements come at a negligible-to-moderate sacrifice in ground user performance compared to scenarios without UAVs.

To the best of our knowledge, this is the first work that determines the necessary conditions and designs iterative algorithms to optimize cellular networks for UAV corridors using quantization theory. Our findings open avenues for further exploration and extensions from multiple standpoints, some of which are listed as follows: (i) Performance metric, optimizing for capacity per user, rather than SINR, thus aligning more closely with the objectives of real-world mobile network operators; 
(ii) Antenna pattern, considering BSs transmitting multiple beamformed synchronization signal blocks (SSBs), instead of a single beam, and addressing the optimization of the SSB codebooks;
(iii) Cellular deployment, exploring the optimization of BS locations, in addition to their antenna tilts and transmit power; and
(iv) Channel model, replacing the statistical 3GPP model with a scenario-specific map-based channel model, providing a more accurate, ad-hoc representation of the channel characteristics. 
Progress along any of the above directions would extend the applicability and scope of our work, paving the way for advancements in optimizing cellular networks for UAV corridors and addressing emerging challenges in air-to-ground wireless communications.
\appendices
\section{Proof of Proposition \ref{optimal-V-SINR}}\label{Appendix_A}

The equivalence between Eqs. (\ref{optimal-cell-partitioning-SINR}) and  (\ref{optimal-cell-partitioning-SINR-proof}) is shown on top of the next page, where
\begin{figure*}[t!]
\begin{align}\label{optimal-cell-partitioning-SINR-proof}
    V_n^*(\bm{\Theta}, \bm{\rho}) &= \big\{\bm{q} \in Q \mid \mathtt{SINR_{dBm}^{(n)}}(\bm{q}; \bm{\Theta}, \bm{\rho}) \geq \mathtt{SINR_{dBm}^{(k)}}(\bm{q}; \bm{\Theta}, \bm{\rho}), \quad \textrm{ for all } 1 \leq k \leq N \big\} \\
    &= \left\{\bm{q} \in Q \,\middle\vert\,  \frac{\mathtt{RSS^{(n)}_{lin}} (\bm{q}; \theta_n, \rho_n)}{\sum\limits_{j\neq n}^{} \mathtt{RSS^{(j)}_{lin}} (\bm{q}; \theta_j, \rho_j) + \sigma^2} \geq \frac{\mathtt{RSS^{(k)}_{lin}} (\bm{q}; \theta_k, \rho_k)}{\sum\limits_{j\neq k}^{} \mathtt{RSS^{(j)}_{lin}} (\bm{q}; \theta_j, \rho_j) + \sigma^2}, \textrm{ } \forall \textrm{ } 1 \leq k \leq N \right\} \nonumber\\
    &= \left\{\bm{q} \in Q \,\middle\vert\, \frac{\mathtt{RSS^{(n)}_{lin}} (\bm{q}; \theta_n, \rho_n)}{\Gamma - \mathtt{RSS^{(n)}_{lin}} (\bm{q}; \theta_n, \rho_n)} \geq \frac{\mathtt{RSS^{(k)}_{lin}} (\bm{q}; \theta_k, \rho_k)}{\Gamma - \mathtt{RSS^{(k)}_{lin}} (\bm{q}; \theta_k, \rho_k)}, \quad \textrm{ for all } 1 \leq k \leq N \right\} \nonumber \\ 
    &= \big\{\bm{q} \in Q \mid \mathtt{RSS_{lin}^{(n)}}(\bm{q}; \theta_n, \rho_n) \geq \mathtt{RSS_{lin}^{(k)}}(\bm{q}; \theta_k, \rho_k), \quad \textrm{ for all } 1 \leq k \leq N \big\} \nonumber \\
    &= \big\{\bm{q} \in Q \mid \mathtt{RSS_{dBm}^{(n)}}(\bm{q}; \theta_n, \rho_n) \geq \mathtt{RSS_{dBm}^{(k)}}(\bm{q}; \theta_k, \rho_k), \quad \textrm{ for all } 1 \leq k \leq N \big\}.\nonumber
\end{align}
\end{figure*}
$\Gamma = \sum_{j=1}^{N} \mathtt{RSS^{(j)}_{lin}} (\bm{q}; \theta_j, \rho_j) + \sigma^2$ and $._{\mathtt{lin}}$ denotes linear units (as opposed to $\mathtt{dBm}$). 
For any arbitrary cell partitioning $\bm{W} = (W_1, \cdots, W_N)$, we can write:
\begin{align}
    &\Phi_{\mathtt{SINR}}(\bm{W},\mathbf{\Theta}, \bm{\rho}) = \sum_{n=1}^{N} \int_{W_n} \mathtt{SINR_{dB}^{(n)}} (\bm{q}; \mathbf{\Theta}, \bm{\rho}) \lambda(\bm{q}) d\bm{q} \\ &\leq \sum_{n=1}^{N} \int_{W_n} \max_k \Big[\mathtt{SINR_{dB}^{(k)}} (\bm{q}; \mathbf{\Theta}, \bm{\rho})\Big] \lambda(\bm{q}) d\bm{q} \nonumber\\
    &=\int_Q \! \max_k \! \Big[\mathtt{SINR_{dB}^{(k)}} (\bm{q}; \mathbf{\Theta}, \bm{\rho})\Big] \lambda(\bm{q}) d\bm{q} \nonumber\\&=\sum_{n=1}^{N} \! \int_{V_n^*} \! \max_k \! \Big[\mathtt{SINR_{dB}^{(k)}} (\bm{q}; \mathbf{\Theta}, \bm{\rho}))\Big] \lambda(\bm{q}) d\bm{q} \nonumber\\&
    \stackrel{(\text{a})}{=} \sum_{n=1}^{N} \int_{V_n^*} \mathtt{SINR_{dB}^{(n)}} (\bm{q}; \mathbf{\Theta}, \bm{\rho}) \lambda(\bm{q}) d\bm{q} = \Phi_{\mathtt{SINR}}(\bm{V}^*, \mathbf{\Theta}, \bm{\rho}),\nonumber
\end{align}
where (a) follows from the definition of $V_n^*$ in Eq. (\ref{optimal-cell-partitioning-SINR}) and its equivalency to Eq. (\ref{optimal-cell-partitioning-SINR-proof}), and the proof is complete. $\hfill\blacksquare$

\section{Proof of Proposition \ref{sinr-power-allocation-gradient}}\label{Appendix_B}

First, we derive the partial derivative of the SINR w.r.t. the BS transmission power $\rho_n$:
\begin{align}
    &\frac{\partial \mathtt{SINR_{dB}^{(n)}}(\bm{q};\bm{\Theta},\bm{\rho})}{\partial \rho_n} = \frac{\partial 10 \log_{10}\bigg(\frac{\mathtt{RSS_{lin}^{(n)}}(\bm{q};\theta_n,\rho_n)}{\sum_{j\neq n}\mathtt{RSS_{lin}^{(j)}}(\bm{q};\theta_j,\rho_j) + \sigma^2 }\bigg) }{\partial \rho_n} \nonumber\\
    &= 10 \log_{10}(e)  
    \times \frac{\bigg(\frac{\sum_{j\neq n}\mathtt{RSS_{lin}^{(j)}}(\bm{q};\theta_j,\rho_j) + \sigma^2 }{\mathtt{RSS_{lin}^{(n)}}(\bm{q};\theta_n,\rho_n)}\bigg)}{\sum_{j\neq n}\mathtt{RSS_{lin}^{(j)}}(\bm{q};\theta_j,\rho_j) + \sigma^2} \nonumber \\ & \times \frac{\partial 10^{\frac{\mathtt{RSS_{dBm}^{(n)}}(\bm{q};\theta_n,\rho_n)}{10}} }{\partial \rho_n} 
    = \frac{10 \log_{10}(e)}{\mathtt{RSS_{lin}^{(n)}}(\bm{q};\theta_n,\rho_n)} \times 10^{\frac{\mathtt{RSS_{dBm}^{(n)}}(\bm{q};\theta_n,\rho_n)}{10}} \nonumber\\& \times \frac{\ln(10)}{10} \times \frac{\partial \mathtt{RSS_{dBm}^{(n)}}(\bm{q};\theta_n,\rho_n)}{\partial \rho_n} 
    = 1, \label{sinr-power-allocation-gradient-proof-eq1}
\end{align}
and for $i\neq n$, we have:
\begin{align}
    &\frac{\partial \mathtt{SINR_{dB}^{(i)}}(\bm{q};\bm{\Theta},\bm{\rho})}{\partial \rho_n} = \frac{\partial 10 \log_{10}\bigg(\frac{\mathtt{RSS_{lin}^{(i)}}(\bm{q};\theta_i,\rho_i)}{\sum_{j\neq i}\mathtt{RSS_{lin}^{(j)}}(\bm{q};\theta_j,\rho_j) + \sigma^2 }\bigg) }{\partial \rho_n} \nonumber\\&
    = -10 \log_{10}(e) \times \bigg(\frac{\sum_{j\neq i}\mathtt{RSS_{lin}^{(j)}}(\bm{q};\theta_j,\rho_j) + \sigma^2 }{\mathtt{RSS_{lin}^{(i)}}(\bm{q};\theta_i,\rho_i)}\bigg) \nonumber \\&
    \times \frac{\mathtt{RSS_{lin}^{(i)}}(\bm{q};\theta_i,\rho_i) \times \mathtt{RSS_{lin}^{(n)}}(\bm{q};\theta_n,\rho_n) \times \frac{\ln(10)}{10}  }{\Big[\sum_{j\neq i}\mathtt{RSS_{lin}^{(j)}}(\bm{q};\theta_j,\rho_j) + \sigma^2 \Big]^2} \nonumber\\& \times \frac{\partial \mathtt{RSS_{dBm}^{(n)}}(\bm{q};\theta_n,\rho_n)}{\partial \rho_n} 
    = - \frac{\mathtt{RSS_{lin}^{(n)}}(\bm{q};\theta_n,\rho_n)}{\sum_{j\neq i}\mathtt{RSS_{lin}^{(j)}}(\bm{q};\theta_j,\rho_j) + \sigma^{2} } \nonumber\\&
    = - \frac{\mathtt{RSS_{lin}^{(n)}}(\bm{q};\theta_n,\rho_n) \times \mathtt{SINR_{lin}^{(i)}}(\bm{q};\bm{\Theta},\bm{\rho})}{\mathtt{RSS_{lin}^{(i)}}(\bm{q};\theta_i,\rho_i)}. \label{sinr-power-allocation-gradient-proof-eq2}
\end{align}
Similar to the proof of Proposition \ref{gradient-equation-SINR}, the partial derivative component corresponding to the integral over the boundary of regions will sum to zero. Hence, we have:
\begin{equation}\label{sinr-power-allocation-gradient-proof-eq3}
    \frac{\partial \Phi_{\mathtt{SINR}}(\bm{V},\mathbf{\Theta}, \bm{\rho})}{\partial \rho_n} = 
    \sum_{i=1}^{N} \int_{V_i(\mathbf{\Theta}, \bm{\rho})} \!\!\!\!\frac{\partial \mathtt{SINR_{dB}^{(i)}} (\bm{q}; \mathbf{\Theta}, \bm{\rho})}{\partial \rho_n}  \lambda(\bm{q})d\bm{q}.
\end{equation}
Eq. (\ref{sinr-power-allocation-gradient-equation}) then follows from substitution of Eqs. (\ref{sinr-power-allocation-gradient-proof-eq1}) and (\ref{sinr-power-allocation-gradient-proof-eq2}) into Eq. (\ref{sinr-power-allocation-gradient-proof-eq3}). $\hfill\blacksquare$

\balance

\bibliographystyle{IEEEtran}
\bibliography{main}

\end{document}